\documentclass[10pt,a4paper,twoside,american]{article}
\usepackage{geometry}
\usepackage{authblk}
\usepackage{calc}
\usepackage{graphicx}
\usepackage[table]{xcolor}
\definecolor{shadecolor}{gray}{0.9}
\usepackage{natbib} 
\usepackage[font=small,labelfont=bf]{caption}
\usepackage{amsmath}
\usepackage{amssymb}
\usepackage{mathtools}
\usepackage{bm}

\usepackage[
breaklinks=true,
colorlinks=true,
linkcolor=gray,citecolor=gray,urlcolor=gray,filecolor=gray,
pdffitwindow,
bookmarks=true,
bookmarksopen=true,
bookmarksnumbered=true,
pdftitle={},
pdfauthor={},
pdfsubject={},
pdfkeywords={}
]
{hyperref}
\usepackage{changes} 
\definecolor{CB}{RGB}{0,0,200}
\definecolor{TT}{RGB}{0,200,0}
\definecolor{RD}{RGB}{200,200,0}
\definecolor{AM}{RGB}{200,0,0}
\definechangesauthor[color=CB]{CB}
\definechangesauthor[color=TT]{TT}
\definechangesauthor[color=RD]{RD}
\definechangesauthor[color=AM]{AM}

\usepackage{cleveref}
\crefformat{equation}{(#1)}
\crefformat{figure}{Fig.\,#1}
\crefformat{section}{Sec.\,#1}
\crefformat{table}{Tab.\,#1}

\usepackage[]{float}
\restylefloat{figure}

\def\figdir{.}

\newcommand*{\synweight}{J}

\newcommand*{\tobs}{t_{\text{obs}}}

\newcommand*{\numstims}{p}

\newcommand*{\firingrate}{\nu}

\newcommand{\CM}{C_\text{m}}    
\newcommand{\CV}{\text{CV}}     
\newcommand{\dtfilter}{\Delta t_\text{f}}
\newcommand{\dtpert}{\delta t^*}
\newcommand{\dtsim}{\Delta t}
\newcommand{\EE}{{\exc\exc}}
\newcommand{\EI}{{\exc\inh}}
\newcommand{\exc}{\text{E}}     
\newcommand{\ext}{\text{X}}   
\renewcommand{\exp}{\text{exp}} 
\newcommand{\erf}{\text{erf}} 
\newcommand{\EW}[2][]{\left\langle{#2}\right\rangle_{#1}}
\newcommand{\FF}{\text{FF}}     
\newcommand{\IE}{{\inh\exc}}
\newcommand{\II}{{\inh\inh}}
\newcommand{\Epop}{\mathcal{E}} 
\newcommand{\inh}{\text{I}}     
\newcommand{\Ipop}{\mathcal{I}} 

\newcommand{\PSCamp}{\hat{I}}               
\newcommand{\J}{J}                          

\newcommand{\JEE}{\J_{\exc\exc}}
\newcommand{\JEI}{\J_{\exc\inh}}

\newcommand{\JIE}{\J_{\inh\exc}}
\newcommand{\JII}{\J_{\inh\inh}}

\newcommand{\Jfb}{\hat{J}}                   

\newcommand{\JfbEE}{\Jfb_{\exc\exc}}
\newcommand{\JfbEI}{\Jfb_{\exc\inh}}

\newcommand{\JfbIE}{\Jfb_{\inh\exc}}
\newcommand{\JfbII}{\Jfb_{\inh\inh}}
\newcommand{\JfbX}{{\Jfb_\ext}}
\newcommand{\K}{K}

\newcommand{\KEE}{{K_{\exc\exc}}}
\newcommand{\KEI}{{K_{\exc\inh}}}

\newcommand{\KIE}{{K_{\inh\exc}}}
\newcommand{\KII}{{K_{\inh\inh}}}
\newcommand{\Kinput}{{K^\text{out}_\ext}} 
\newcommand{\KX}{{K_\ext}} 
\newcommand{\mat}[1]{\bm{#1}}
\renewcommand{\max}{\text{max}} 
\newcommand{\ms}{\,\text{ms}}
\newcommand{\muE}{{\mu_\exc}}
\newcommand{\muI}{{\mu_\inh}}
\newcommand{\mV}{\,\text{mV}}
\newcommand{\NE}{{N_{\exc}}}
\newcommand{\NI}{{N_{\inh}}}
\newcommand{\nuE}{\nu_\exc}
\newcommand{\nuI}{\nu_\inh}
\newcommand{\nuX}{\nu_\ext}

\newcommand{\pF}{\,\text{pF}}
\newcommand{\Real}{\text{Re}}
\newcommand{\RM}{R_\text{m}}
\newcommand{\SD}{\text{SD}}
\newcommand{\seconds}{\,\text{s}}
\newcommand{\sigmaE}{\sigma_{\exc}}
\newcommand{\sigmaI}{\sigma_{\inh}}
\newcommand{\sps}{\,\text{spikes/s}}
\newcommand{\taufilter}{\tau_\text{f}}
\newcommand{\tauM}{\tau_\text{m}}
\newcommand{\tauR}{\tau_\text{ref}}
\newcommand{\tauS}{\tau_\text{s}}
\newcommand{\transp}{^\text{T}}
\newcommand*{\Var}{\text{Var}}
\renewcommand{\vec}{\bm}
\newcommand{\Vreset}{V_\text{r}}
\newcommand{\effweight}{w}                   

\newcommand{\wEE}{\effweight_{\exc\exc}}
\newcommand{\wEI}{\effweight_{\exc\inh}}

\newcommand{\wIE}{\effweight_{\inh\exc}}
\newcommand{\wII}{\effweight_{\inh\inh}}
\newcommand{\yreset}{y_\text{r}}
\newcommand{\ytheta}{y_\theta}

\title{Firing rate homeostasis counteracts changes in stability of recurrent neural networks caused by synapse loss in Alzheimer's disease}

\author[1,*]{Claudia Bachmann}
\author[1]{Tom Tetzlaff} 
\author[1]{Renato Duarte}
\author[1,2]{Abigail Morrison}
\affil[1]{\footnotesize%
  Institute of Neuroscience and Medicine (INM-6) and Institute for Advanced Simulation (IAS-6) and JARA BRAIN Institute I, J\"ulich Research Centre, J\"ulich, Germany}
\affil[2]{Institute of Cognitive Neuroscience, Faculty of Psychology, Ruhr-University Bochum, Bochum, Germany}
\affil[*]{c.bachmann@fz-juelich.de}

\date{\footnotesize\today}

\begin{document}

\maketitle

\abstract{
The impairment of cognitive function in Alzheimer's is clearly correlated to synapse loss.  However, the mechanisms underlying this correlation are only poorly understood. Here, we investigate how the loss of excitatory synapses in sparsely connected random networks of spiking excitatory and inhibitory neurons alters their dynamical characteristics. Beyond the effects on the network's activity statistics, we find that the loss of excitatory synapses on excitatory neurons shifts the network dynamic towards the stable regime. The decrease in sensitivity to small perturbations to time varying input can be considered as an indication of a reduction of computational capacity.  A full recovery of the network performance can be achieved by firing rate homeostasis, here implemented by an up-scaling of the remaining excitatory-excitatory synapses. By analysing the stability of the linearized network dynamics, we explain how homeostasis can simultaneously maintain the network's firing rate and sensitivity to small perturbations.}

\section*{Author summary}
Relating the properties of neuronal circuits with concrete functional roles pertaining to cognition and behavior is a complex endeavour. This is especially true when it comes to diseases and dysfunctions, where we are often left with high-level clinical observations (e.g. cognitive deficits and structural brain changes), without understanding their relationship. A potentially fruitful approach to address this problem consists of employing simplified mathematical models to test the relevant hypotheses, incorporating pathophysiological observations and evaluating their tentative functional consequences, in relation to clinical observations. In this work, we employ a spiking neural network model to study the effects of synaptic loss, as it is often observed in various neurodegenerative disorders, in particular Alzheimer's disease (AD). We show that the loss of synapses drives the network into a less sensitive regime, which potentially accounts for the cognitive deficits of AD. We also endow the circuits with a compensation mechanism which increases the weight of remaining connections to restore the mean network activity. We demonstrate that this very simple compensatory mechanism can recover all the dynamical features that are changed due to synapse loss. We further develop an analytical model that accounts for this surprising finding.

\section{Introduction}

Accelerated synapse loss is a prominent feature in many types of neurodegenerative disorders, such as Huntington's disease, frontotemporal dementia or Alzheimer's disease \citep{Zhan1993,Brun1995,Morton2001,Lin02_449,Scheff2014}. In Alzheimer's disease (AD), synapse loss appears to be particularly important, as it is widespread across different brain areas and constitutes a key marker in the AD pathology \citep[see e.g.][]{Scheff2014}.

The mechanisms underlying AD related synaptic modifications are currently the subject of intensive research, which has revealed that a number of different alterations at the molecular level may ultimately lead to synaptic decay \citep{Sheng2012,Dorostkar2015,Tampellini2015}, such as an abnormal occurrence of oligomeric and aggregated $\beta$-amyloid-peptides (A$\beta$), an abnormal phosphorylation of the tau protein
and the occurence of neurofibrillary tangles, and a disrupted signaling in neuroinflammatory and oxidative stress responses \citep{Tampellini2015,Toennies2017,Frere2017,Rajendran2018}.

\par
Previous studies have uncovered a strong positive correlation between cognitive impairment in AD patients and synapse loss \citep{DeKosky90,Scheff1990,Terry91_572,Scheff1993,Masliah1994,Scheff03_1029,Scheff2006,Scheff2011}. In contrast, correlations between the cognitive status and the density of plaques or tangles have frequently been reported as rather weak.
Synapse loss is therefore not merely a structural epiphenomenon of AD, but appears to be \emph{the} physical correlate of cognitive decline.

While the most commonly reported early symptom of AD is memory deterioration, the disease is associated with a wide range of other cognitive problems such as stereotyped, repetitive linguistic production, visuo-spatial deficits and disorientation, apraxia, and loss of executive functions, i.e.~planning and abstract reasoning \citep{Bennett2002,Weintraub2012}.
The observed progression of cognitive symptoms goes hand in hand with brain tissue atrophy \citep{Smith2002,ToledoMorrell2000,Thompson2003} associated with loss of synapses \citep{Chen2018}, suggesting that the synaptic degeneration may underlie the cognitive deterioration following the gradual involvement of different, functionally specialized brain regions.

\par

It is known that AD-related molecular and cellular alterations, such as abnormal depositions of A$\beta$ plaques or atrophy rates, often significantly precede cognitive symptoms (see, e.g., \citealp{Sperling2014,Jack10_119}, and references therein).

However, mechanisms exist that counteract synapse loss \citep{Small2004,Fernandes2016}, at least in the early stages of the disease.
Various studies have shown that the loss of synapses is accompanied by a growth of remaining synapses, such that the total synaptic contact area (TSCA) per unit volume of brain tissue is approximately preserved \citep{DeKosky90, Scheff03_1029,Scheff2006,Neuman2014}.
It is likely that such compensatory mechanisms underlie the observed delay in the onset of cognitive symptoms with respect to the onset of symptoms at the cellular level \citep{Morris2005}.

The heterogeneity in the disease progression and the propensity to transition from healthy cognitive aging to mild cognitive impairment and dementia may thus be associated to a subject's ability to counteract synapse loss and, to a certain extent, maintain global functionality in a way that masks the progressive underlying pathophysiology.
Such homeostatic, regulatory mechanisms appear to play an important role in counteracting structural deterioration and preserving computational capabilities.
On the other hand, they pose important challenges to the network's functionality since they have the potential to disrupt the specificities of a circuit's microconnectivity (namely the distribution of synaptic strengths) and thus degrade its information content (e.g. \citealt{Fröhlich2008}).
Successful homeostatic compensation thus requires a balanced orchestration which preserves the system's computational properties and macroscopic dynamics, e.g., average firing rates \citep{Luetcke2013,Slomowitz2015} and E/I balance \citep{Zhou2018}, as well as the relative ratios and distributions of synaptic strengths (e.g. synaptic scaling mechanisms; \citealt{Keck2013,Vitureira2013}).

Understanding the circuit-level consequences of synaptic alterations, entailing both the deregulation by synapse loss and recovery through homeostasis, is essential to understand whether they represent a negative symptom of the disease or a compensatory response. One likely effect is the modification of the network's firing rate. In order to maintain a physiological operating regime far from activity extremes (quiescence or epileptic activity), a network needs the capacity to regulate its firing rate.

The degree to which this may be impaired in AD is still under debate \citep[see][]{Styr2018,Frere2017}. Whereas the effects of synaptic alterations on the network dynamics have been partially characterized, a direct link between synapse loss, network dynamics and functional decline has yet to be systematically established, with only a few studies addressing the topic \citep{Horn1993,Horn1996,Ruppin1994}. However, this connection may prove fruitful, both for understanding the disease itself and for fostering the development of new diagnostic and therapeutic approaches. 

It is currently unknown to what extent homeostatic mechanisms, such as increasing the synaptic area \citep{DeKosky90,Scheff2006}, can completely recover the neuronal network's firing rate, nor whether the preservation of the firing rate by such mechanisms entails the preservation of cognitive performance. In this study, we investigate the link between structure, dynamics and function using a recurrent spiking neural network model \citep{Brunel99}.

Despite their simplicity, such systems have been shown to support computations, such as e.g. stimulus categorization, associative learning and memory, information routing and propagation, etc. \citep[see, e.g.][]{Jaeger04_87,Eliasmith02,Maass02_2531,Buesing11,Boerlin13,Abbott16_350}. Additionally, although these models have complex behavioral repertoires, they are often simple enough that their dynamics can be assessed analytically. The stability of the dynamics can then be related to computational task performance, such as the network's sensitivity to perturbation and classification capability \citep{Legenstein07_323,Legenstein07_127}. Thus, an analytical treatment of network dynamics can provide insight into why some realizations of such networks perform better than others and how performance is affected by structural changes. Theoretical studies explicitly addressing this issue have so far focused either on the disruption of oscillations or functional connectivity of the whole brain, or on memory only (especially memory retrieval; \citealp{Horn1993,Horn1996,Ruppin1994}).

Here, we investigate how the loss of excitatory-excitatory synapses in sparsely connected random networks of spiking excitatory and inhibitory neurons (\cref{subsec:Computational-model-AD}) and firing rate homeostasis, based on upscaling the remaining excitatory-excitatory connections, alters the dynamical characteristics of a network. Surprisingly, we find that firing rate homeostasis can restore a variety of dynamic features caused by synaptic loss, including the increase in spike train regularity, the drop in the fluctuations of population activity and the reduction of the synaptic contact area (\cref{subsec:activity_statistics}) caused by synaptic loss. In addition, we observe that synaptic loss decreases the network's sensitivity to small perturbations (\cref{sec:Results-Sensitivity to perturbation}), such that a network operating near the 'edge of chaos' would be shifted by synaptic loss to a more stable regime; a shift which has been shown in previous studies to result in a decrease in computational capacity \citep{Langton90_12,Legenstein07_127,Legenstein07_323,Schrauwen09_1425,Dambre12_514,Schuecker17_arxiv_v3}, and may account for the cognitive deficits observed in Alzheimer's disease. Here, too, firing rate homeostasis counteracts the shift towards the stable regime. We further show that these compensatory mechanisms ultimately become exhausted if physiological limits are placed on the growth of the synapse. As it is not obvious why simply maintaining the firing rate also maintains the stability of the network, we analyze the stability of the linearized network dynamics and discover a strictly monotonic relationship between the firing rate and the spectral radius of the network, which explains the restoration of the dynamics under the influence of firing rate homeostasis (\cref{sec:Results-Sensitivity to perturbation}).

\section{Results}

\subsection{Computational network model of Alzheimer's disease}
\label{subsec:Computational-model-AD}

\begin{figure}
  \centering
\includegraphics[width=\textwidth]{\figdir/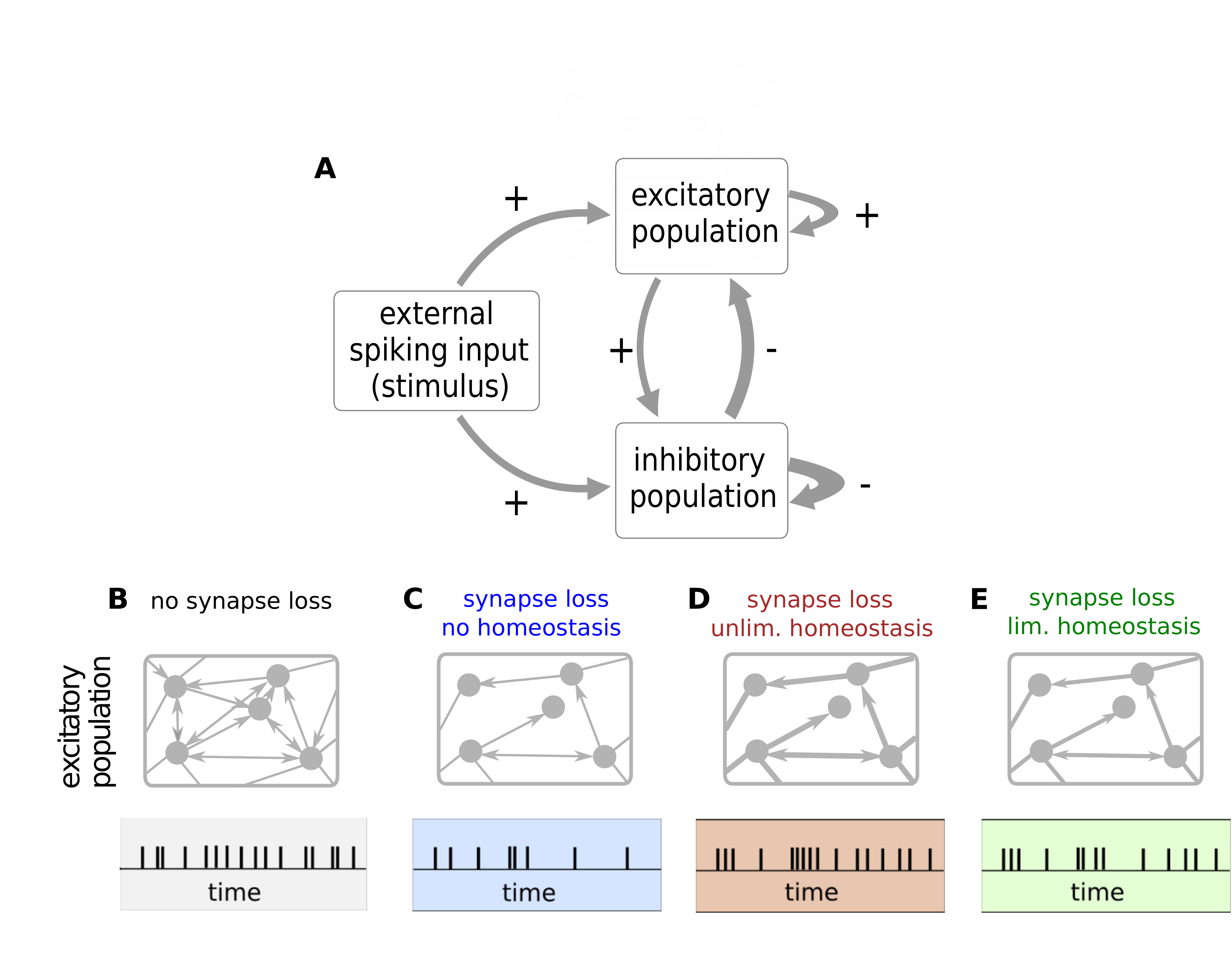}
\caption{%
  \textbf{Sketch of the network model of Alzheimer's disease and homeostasis.} 
  \textbf{A)}
  The network comprises two reciprocally and recurrently connected populations of excitatory (E) and inhibitory (I) integrate-and-fire neurons, excited by an external spiking input. Thickness of arrow indicates relative strength of the connection. In this study, Alzheimer's disease is modeled by removing connections between excitatory neurons (loss of $\EE$ synapses) and upscaling of the remaining $\EE$ synapses to maintain the average firing rate (firing rate homeostasis).
  \textbf{B}--\textbf{E)} Sketch of $\EE$ connection density (number of arrows in upper panels), connection strength (thickness of arrows in upper panels) and resulting single-neuron spiking activity (lower panels).
  \textbf{B)} Intact network (without synapse loss). 
  \textbf{C)} Synapse loss without homeostasis: removal of $\EE$ synapses and resulting reduction in firing rate. 
  \textbf{D)} Synapse loss  with unlimited homeostasis: removal of $\EE$ synapses and increase in strength of remaining $\EE$ synapses to maintain the average firing rate. Synaptic weights are allowed to grow without bounds.
  \textbf{E)} Synapse loss  with limited homeostasis: removal of $\EE$ synapses and bounded increase in strength of remaining $\EE$ synapses. Here, synaptic weights cannot exceed $120\%$ of their reference weight. The firing rate is therefore only partially recovered.
  For a complete description and parameter specification of the network model, see \cref{subsec:network_modelling} and Supplementary Material (\cref{sec:suppl_network_model}, \cref{sec:suppl_parameters}).
}
\label{fig:The-computational-model}
\end{figure}

We study the effects of AD related synaptic alterations on the network dynamics and computational characteristics in the framework of a generic mathematical neuronal network model (\cref{fig:The-computational-model}\,A), which captures prominent structural and dynamical features of local neocortical networks such as the relative numbers of excitatory and inhibitory neurons \citep{Scholl1956,Abeles82} and  synapses \citep{DeFelipe1992, Gulyas1999}, sparse connectivity \citep{Abeles82,Binzegger04}, small synaptic weights \citep{Lefort2009}, irregular \citep{Tomko1974,Softky93,Shadlen98} and predominantly asynchronous spiking \citep{Ecker10}, large membrane potential fluctuations \citep{Petersen13,Cowan94_17,Timofeev2001_1924,Steriade93}, and a tight dynamical balance between excitatory and inhibitory synaptic currents \citep{Okun2008_535}.

The network is composed of randomly and sparsely connected populations of excitatory ($\exc$) and inhibitory ($\inh$) integrate-and-fire neurons, driven by external spiking input.
The overall coupling strength is determined by the reference synaptic weight $J$. For simplicity, all excitatory connections ($\EE$ and $\IE$) and all inhibitory connections ($\EI$ and $\II$), respectively, have equal synaptic weight:~$\JEE=\JIE=J$ and $\JEI=\JII=-gJ$ in the intact network (i.e. before synapse loss). The relative strength $g$ of inhibitory weights is chosen such that the network is dominated by inhibition, to permit asynchronous irregular firing at low rates \citep{Brunel00}.
A complete specification of the network model and parameters can be found in \cref{subsec:network_modelling} and in the Supplementary Material (\cref{sec:suppl_network_model}, \cref{sec:suppl_parameters}). An illustration of the connectivity of the excitatory population for an intact network and an example spike train is given in  \cref{fig:The-computational-model}\,B.

We implement the effects of AD on the network connectivity by reducing the number of excitatory synapses on excitatory neurons ($\EE$ synapses; \citealp{Lacor07_796,Dorostkar2015}), whilst keeping the number of connections between other populations (EI, IE, II) constant. In the absence of any compensation mechanism, this modification leads to a reduction in the average firing rate (see \cref{subsec:activity_statistics}).

In biological neuronal networks, long-term activity levels are often stabilized by homeostatic regulation \citep{Bennett2002,Marder06_563,Turrigiano2008}.
While a maintenance  of firing rates has been observed at the level of individual neurons \citep{Luetcke2013}, long-term recordings suggest a predominance of a network-wide regulation \citep{ Slomowitz2015} targeting a constant population firing rate.
Such a homeostatic stabilization of the population firing rate can be accounted for by a global adjustment of synaptic weights  (synaptic scaling; \citealp{Vitureira2012,Turrigiano2012review}). 
Indeed, in the early stages of AD, synapse loss seems to be compensated by a growth of the remaining synapses \citep{DeKosky90, Scheff2006,Neuman2014}.  
To realize this mechanism in our spiking neuronal network, we implement a firing-rate homeostasis which compensates for the loss of $\EE$ synapses by a global increase in the weights $\JEE$ of the remaining  $\EE$ synapses, thereby preserving the population firing rate. 

For advanced AD, where a large portion of the $\EE$ synapses has been lost, a full recovery of the population firing rate through synaptic scaling would require unrealistically large synaptic weights. 
During aging and dementia, the maximum increase in synaptic size has been reported to be in the range from $9\%$ to $24\%$ (see \citealp{Scheff03_1029}, and references therein).
We incorporate these findings by introducing an optional upper bound for the weight $\JEE$ of $\EE$ synapses.

To uncover the differential effects of excitatory synapse loss and homeostasis, in this study we investigate the  dynamical and computational characteristics of a network for three different scenarios: 
synapse loss without homeostatic compensation (\cref{fig:The-computational-model}\,C), 
synapse loss with an unlimited firing rate homeostasis where synaptic weights can grow without bounds (\cref{fig:The-computational-model}\,D), and
synapse loss with limited firing rate homeostasis where the synaptic weights cannot exceed $120\%$ of the weight in the intact reference network (\cref{fig:The-computational-model}\,E).

Note that the model's high level of abstraction enables us to identify fundamental mechanisms, to reduce the risk of overfitting, and to arrive at general conclusions that may be transferred to other brain regions or even different spatial scales. Empirically observed features of biological neural networks
such as heavy-tail synaptic weight distributions \citep{Song05_0507,Ikegaya13_293} or active dendritic processing \citep{Major13} are not explicitly incorporated. As a consequence, model parameters such as synaptic weights have to be regarded es ``effective'' parameters and cannot be mapped to biological parameters in a one-to-one fashion. Selecting a particular set of parameters to be considered "biologically realistic" would be misleading. Therefore, rather than focusing on a specific configuration of the model, we systematically vary both the reference synaptic weight $J$ and the extent of synapse loss to uncover the general relationship between these parameters and the dynamical and computational properties of the network.

\subsection{Total synaptic contact area and firing statistics}
\label{subsec:activity_statistics}
\begin{figure}
  \centering
  \includegraphics[width=\textwidth]{\figdir/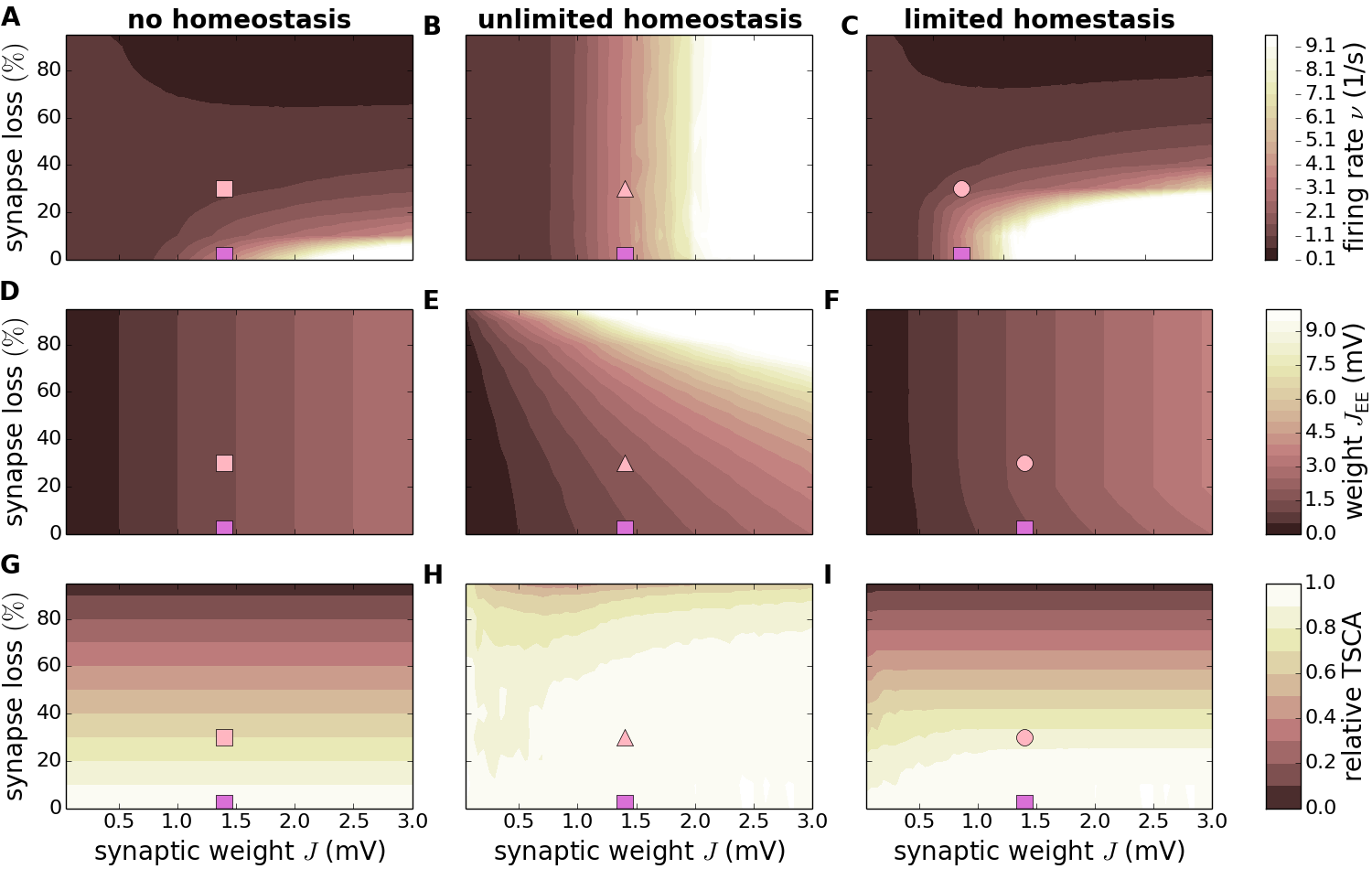}
  \caption{%
    \textbf{Effect of synapse loss and firing rate homeostasis on firing rate, synaptic weights and total synaptic contact area.}
  Dependence of the time and population averaged firing rate $\nu$ (\textbf{A}--\textbf{C}),  synaptic weight $\JEE$ (\textbf{D}--\textbf{F}) and the relative total synaptic contact area (TSCA) of $\EE$ synapses (\textbf{G}--\textbf{I}) on the reference weight $J$ and the degree of $\EE$ synapse loss in the absence of homeostatic compensation (left column),
  as well as with unlimited (middle column) and limited firing rate homeostasis (right column).
  Color-coded data represent mean across $10$ random network realizations.
  Symbols mark parameter configurations shown in \cref{fig:rastaplots}.\
}
  \label{fig:FR-J-plots}
\end{figure}

In the absence of homeostatic compensation (left column of \cref{fig:FR-J-plots}), removal of excitatory synapses on excitatory neurons naturally results in a decrease in the population firing rate $\nu$, irrespective of the synaptic-weight scale $J$ (\cref{fig:FR-J-plots}A).
An upscaling of the remaining $\EE$ synapses (middle column) allows us to preserve the population firing rate, even if substantial amounts of synapses are removed (vertical contours in \cref{fig:FR-J-plots}B).
If the maximum synaptic weight is limited, firing rates are preserved only up to a critical level of synapse loss (early stages of AD; \cref{fig:FR-J-plots}C).
\par
Experimental studies have shown that, in early AD, the reduction in the number of synapses is accompanied by a growth of the remaining synapses such that the total synaptic contact area (TSCA) per unit volume is approximately preserved \citep{DeKosky90, Scheff2006,Neuman2014}.
Our simple AD network model reproduces this finding if we define the TSCA as the product of the number of $\EE$ connections and the synaptic weight $\JEE$ (\cref{subsec:activity_metrics}).
Without homeostatic upscaling of $\EE$ weights, the TSCA is proportional to the number of $\EE$ connections and therefore quickly decreases with increasing levels of synapse loss (\cref{fig:FR-J-plots}G).
In the presence of firing rate homeostasis, however, the TSCA remains largely constant unless a majority of synapses is lost (\cref{fig:FR-J-plots}H) or the maximum synaptic weight is reached (\cref{fig:FR-J-plots}I).
We conclude that the experimentally observed stabilization of the TSCA in the face of synapse loss may be a consequence of a homeostatic synaptic scaling regulated by the average population firing rate.
\par
\begin{figure}
  \centering
\includegraphics[width=\textwidth]{\figdir/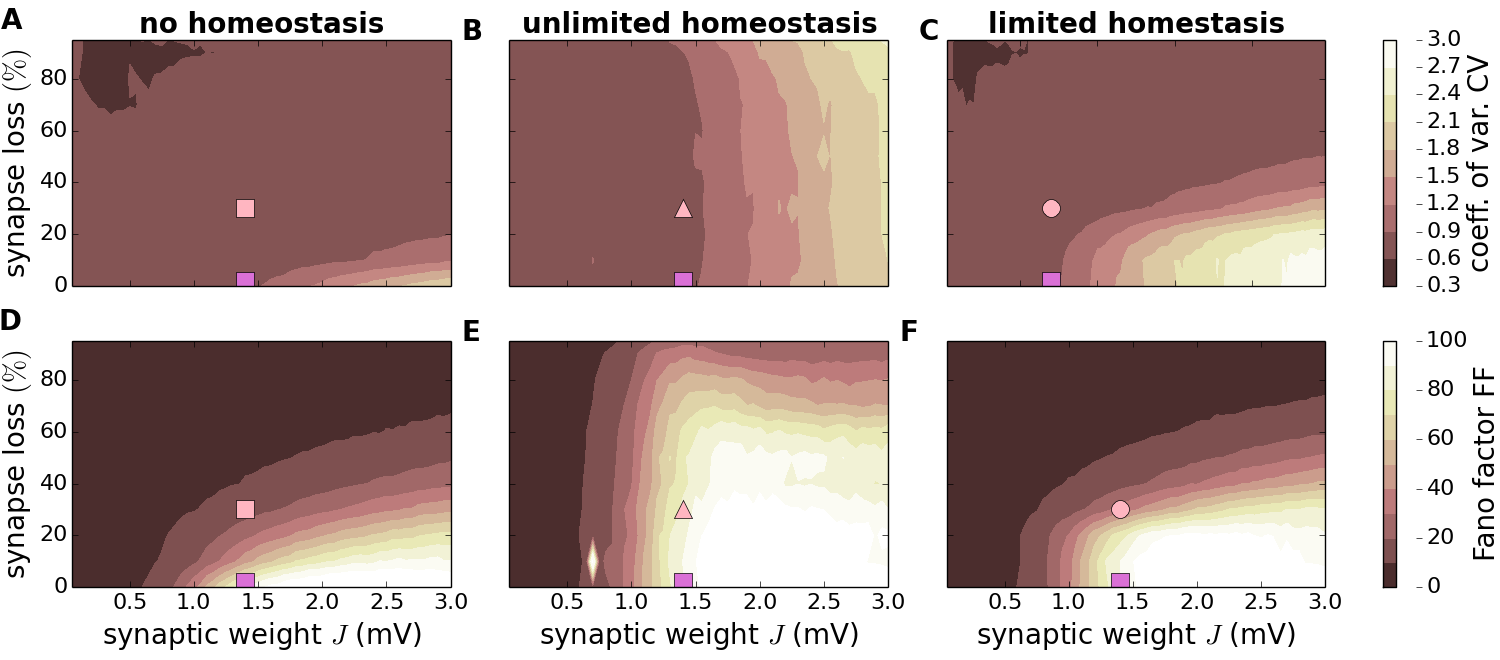}
\caption{%
  \textbf{Effect of synapse loss and firing rate homeostasis on spike train statistics.} 
  Dependence of the coefficient of variation $\CV$ of inter-spike intervals (\textbf{A}--\textbf{C}) and the Fano factor $\FF$ of the population spike count (binsize $b=10\ms$; \textbf{D}--\textbf{F}) on the synaptic reference weight $J$ and the degree of $\EE$ synapse loss in the absence of homeostatic compensation (left column), as well as with unlimited (middle column) and limited firing rate homeostasis (right column).
  Color-coded data represent mean across $10$ random network realizations.
  Symbols mark parameter configurations shown in \cref{fig:rastaplots}.
}
\label{fig:statsplots}
\end{figure}

In physiologically relevant low activity regimes, neuronal firing is determined both by the mean as well as by fluctuations in the synaptic input.
A reduction in the number of synapses followed by an upscaling of synaptic weights may preserve the average population firing rate; it cannot, however, simultaneously preserve the mean and the variance of the synaptic input currents.
The neurons' working point, i.e.~the statistics of the synaptic input, will inevitably change.
A priori, it is therefore not clear to what extent synapse loss and firing rate homeostasis alter the overall firing statistics in the recurrent network beyond the average firing rate.
Here, we address this question by studying the irregularity of spike generation by individual neurons, measured by the coefficient of variation $\CV$ of the inter-spike interval distribution, and spike-train synchrony, assessed by the normalized variance of the population spike count, the Fano factor $\FF$, in $10\ms$ time intervals (see \cref{subsec:activity_metrics}).
Without homeostatic compensation, synapse loss generally results in spike patterns that are less irregular (\cref{fig:statsplots}A) and less synchronous (\cref{fig:statsplots}D).
In the presence of firing rate homeostasis, however, both the $\CV$ and the $\FF$ are largely preserved  (\cref{fig:statsplots}B,E).
Only if the level of synapse loss becomes too severe or if the synaptic-strength limits are reached (limited homeostasis), the $\CV$ and the $\FF$ are reduced (\cref{fig:statsplots}B,C and \cref{fig:statsplots}E,F).

\begin{figure}
  \centering
  \includegraphics[width=\textwidth]{\figdir/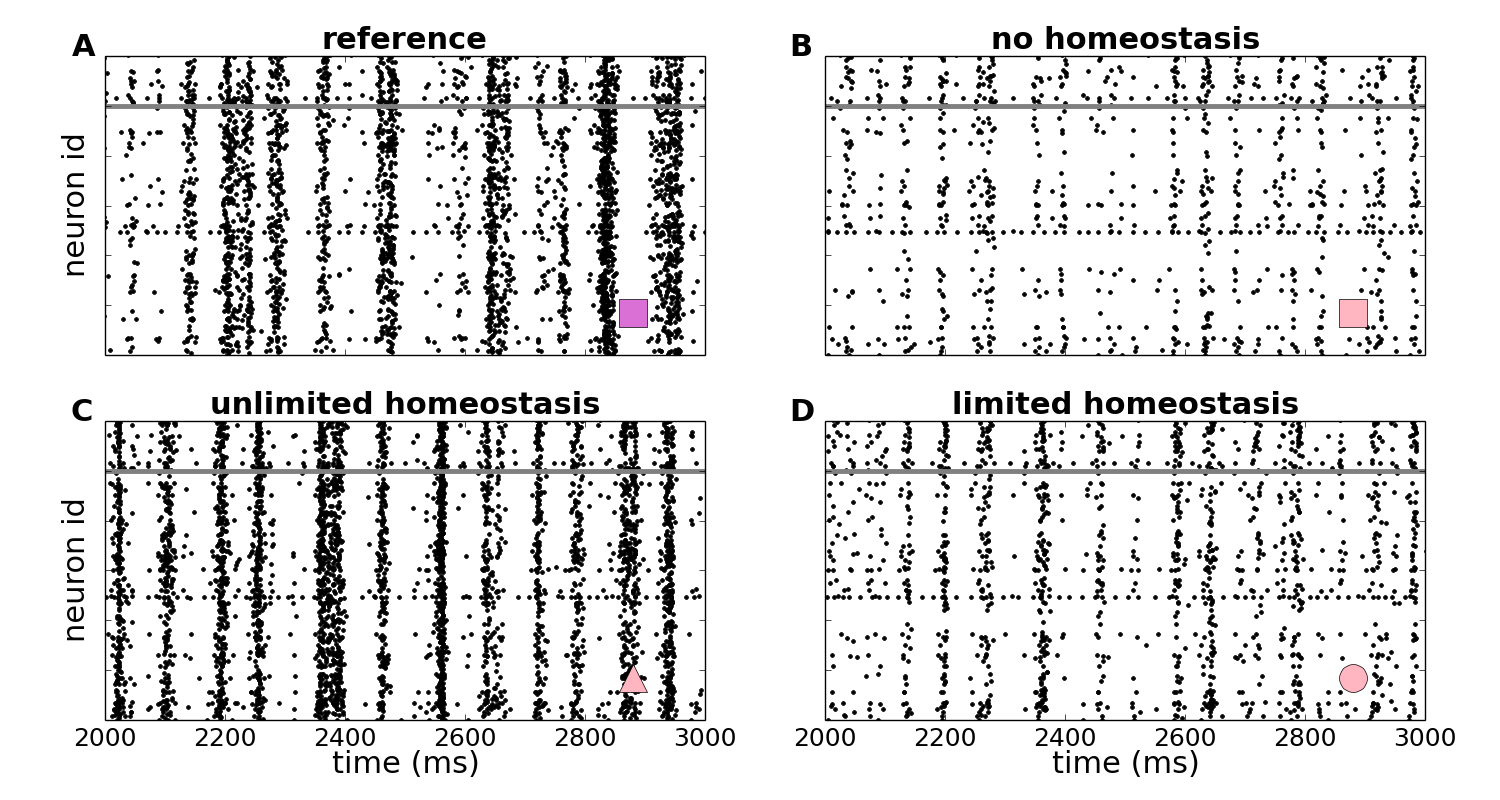}
  \caption{%
    \textbf{Effect of synapse loss and firing rate homeostasis on spiking activity.} 
    Spiking activity (dots mark time and sender of each spike) in
    an intact reference network (no synapse loss, $\JEE=1.4\mV$; \textbf{A}),
    as well as in networks where $30\%$ of the $\EE$ synapses are removed:
    \textbf{B)} no homeostasis ($\JEE=1.4\mV$),
    \textbf{C)} unlimited homeostasis ($\JEE=2.02\mV$),
    \textbf{D)} limited homeostasis ($\JEE=1.68\mV$).
    In all panels, the synaptic-weight scale is set to $\synweight=1.4\mV$.
    Examples depict parameter configurations marked by corresponding symbols in \cref{fig:FR-J-plots} and \cref{fig:statsplots} (cf.~marker in lower right corner of each panel).
    Regions below and above the gray horizontal line show spiking activity of a subset of $100$ excitatory and $25$ inhibitory neurons, respectively.
  }
  \label{fig:rastaplots}
\end{figure}

For illustration, \cref{fig:rastaplots} depicts the spiking activity for four example parameter settings marked by the symbols in \cref{fig:FR-J-plots} and \cref{fig:statsplots}. As  \cref{fig:statsplots}\,E and H already suggest, the overall spiking activity, e.g the number and the duration of synchronous event and the spiking frequency of single neurons, of the homeostatic network (\cref{fig:rastaplots}\,C) and the reference network   (\cref{fig:rastaplots}\,A) are very similar. Only the exact timing of  the synchronous events and the  single neuron spiking differ.
In the AD network without homeostasis (\cref{fig:rastaplots}\,B), the firing rates of both excitatory and inhibitory neurons are decreased. The number of synchronous events, compared with the reference network (\cref{fig:rastaplots}\,A), does not seem to be decreased, but their duration does. The  network with limited homeostasis (\cref{fig:rastaplots}\,D) is more  similar to the AD network without homeostasis than the unlimited homeostasis networks, because the restriction in synaptic growth prevents the rate from being recovered. 

\subsection{Perturbation sensitivity and linear stability}
\label{sec:Results-Sensitivity to perturbation}
\begin{figure}[ht!]
  \centering
  \includegraphics[width=\textwidth]{\figdir/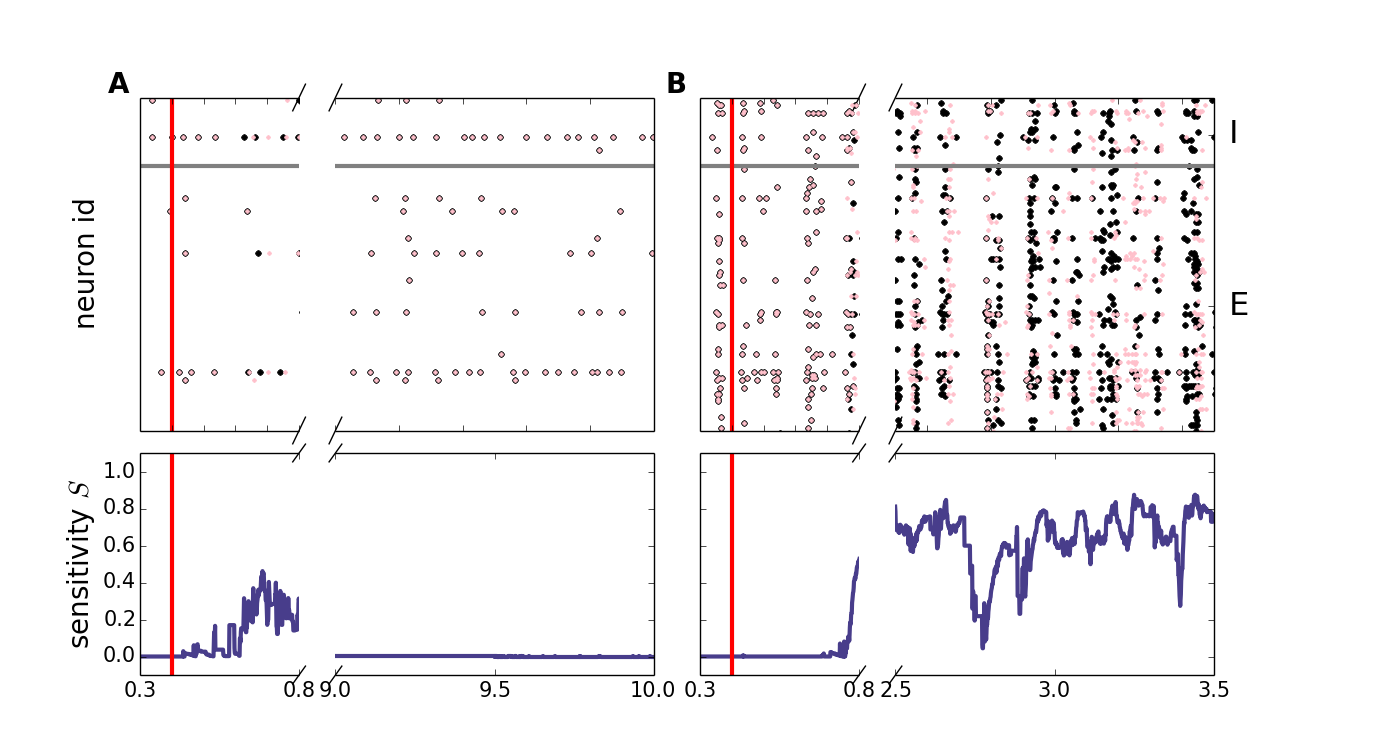}
  \caption{%
    \textbf{Perturbation sensitivity.}
    Top: Example spiking activity (dots mark time and sender of each spike) of two identical networks (identical neuron parameters, connectivity, external input, initial conditions) with (black dots) and without perturbation (purple dots). 
    The perturbation consists in delaying one external input spike  at time $t^*=400\,\ms$ by $\dtpert=0.5\,\ms$.
    The vertical red line marks the time of the perturbation.
    Spikes of only $10\%$ of all neurons are shown.
    Neurons below and above the horizontal gray line correspond to excitatory and inhibitory neurons, respectively.
    Bottom row: Perturbation sensitivity $S(t)=1-|R(t)|$ obtained from the correlation coefficient $R(t)$ of the low-pass filtered spike trains generated by the unperturbed and the perturbed network (black and purple dots in top panels; see Methods \cref{subsec:activity_metrics}).
    \textbf{A)} Stable dynamics ($\J=0.45\mV$, $\KEE=100$).
    \textbf{B)} Chaotic dynamics ($\J=1.75\mV$, $\KEE=100$).    
    Note different time scales in A and B.
  }
    \label{fig:perturbation_sensitivity_measure} 
\end{figure}

An open, yet chronically ignored, question in Alzheimer's disease research is how cellular damage such as synapse loss affects patients' cognitive capabilities.
A number of theoretical studies have shown that recurrent neuronal networks exhibit optimal computational performance characteristics for a variety of task modalities if they operate in a dynamical regime where small perturbations are neither instantly forgotten nor lead to entirely different network states \citep{Langton90_12,Legenstein07_127,Legenstein07_323,Schrauwen09_1425,Dambre12_514,Schuecker17_arxiv_v3}.
In dynamical systems theory, this regime has been termed the ``edge of chaos'' as it represents the transition from a stable state with a low sensitivity to small perturbations to a chaotic state where the sensitivity to small perturbations is high. Here, we investigate the role of synapse loss and firing rate homeostasis for the network's sensitivity to perturbations as an indicator of its overall computational performance.
\par
To assess the perturbation sensitivity, we simulate a given network twice with identical initial conditions and identical realizations of external inputs.
In the second run, we apply a small perturbation by delaying one of the external input spikes to a single neuron by a fraction of a millisecond (\cref{fig:perturbation_sensitivity_measure}).
In stable regimes, the effect of this perturbation on the spiking response is transient and quickly vanishes (\cref{fig:perturbation_sensitivity_measure}\,A, top).
In chaotic regimes, in contrast, the small perturbation leads to diverging spike patterns (\cref{fig:perturbation_sensitivity_measure}\,B, top). 
We quantify the network's perturbation sensitivity $S=1-|R|$ in terms of the long-term correlation coefficient $R$ between the low-pass filtered spike responses in the two runs (\cref{fig:perturbation_sensitivity_measure}, bottom).
With this definition, $S=0$ and $S=1$ correspond to insensitive (stable) and highly sensitive (chaotic) networks, respectively (for details, see \cref{subsec:activity_metrics}).
\par
\begin{figure}[ht!]
  \centering
  \includegraphics[width=\textwidth]{\figdir/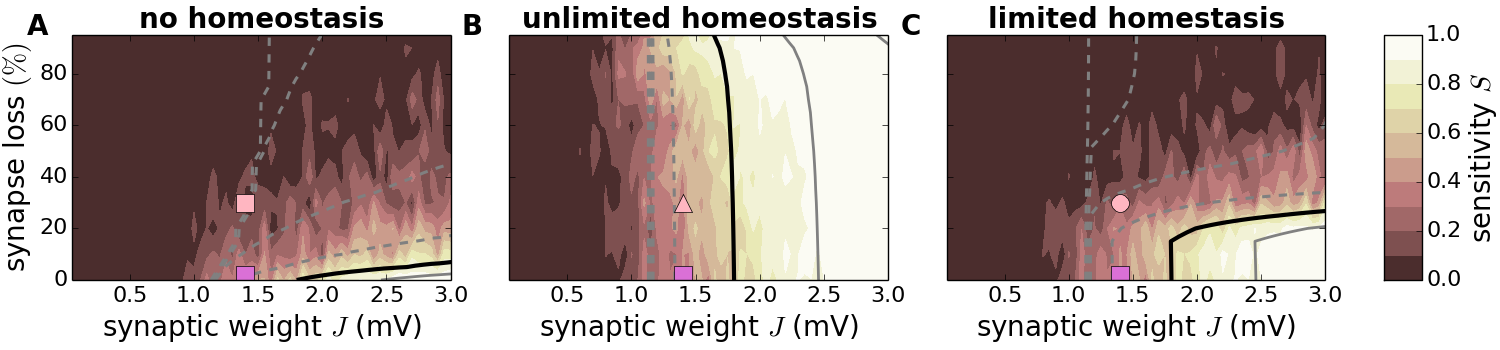}
  \caption{%
    \textbf{Effect of synapse loss and firing rate homeostasis on perturbation sensitivity.}
    Dependence of perturbation sensitivity $S$ 
    on the synaptic reference weight $J$ and the degree of $\EE$ synapse loss
    in the absence of homeostatic compensation (\textbf{A}), as well as for unlimited (\textbf{B}) and limited firing rate homeostasis (\textbf{C}).
    Color-coded data represent mean across $10$ random network realizations.
    Superimposed black and gray curves mark regions where the linearized network dynamics is 
    stable (gray dashed; spectral radius $\rho=\ldots,0.6,0.8$),
    about to become unstable (black; $\rho = 1$), and
    unstable (gray solid; $\rho=1.2,1.4,\ldots$).
    Pink symbols mark parameter configurations shown in \cref{fig:FR-J-plots} and \cref{fig:rastaplots}.
 }
  \label{fig:Sensitivity-to-perturbation}
\end{figure}

For small synaptic weights $J$, the network dynamics is always stable ($S=0$) for our choice of parameters, irrespective of the degree of synapse loss and the absence or presence of homeostatic compensation (\cref{fig:Sensitivity-to-perturbation}).
In this regime, the perturbation has no long-term effect: after a transient phase, the response spike patterns in the perturbed and the unperturbed simulation are exactly identical (at the temporal resolution $\dtfilter=1\ms$ of the recorded signals).
The intact networks (zero synapse loss) enter a chaotic regime ($S>0$) if the synaptic weights $J$ exceed a certain critical value.
Removal of $\EE$-synapses without homeostatic compensation leads to a shift of this transition towards larger synaptic weights (\cref{fig:Sensitivity-to-perturbation}\,A).
Networks in the chaotic regime eventually become insensitive to perturbations with progressing $\EE$-synapse loss.
In the presence of firing rate homeostasis, in contrast, the perturbation sensitivity is preserved (color gradient in \cref{fig:Sensitivity-to-perturbation}\,B is predominantly left to right, rather than top to bottom).
Unless the homeostatic strengthening of $\EE$-synapses is limited (limited homeostasis; \cref{fig:Sensitivity-to-perturbation}\,C), this maintenance of the perturbation sensitivity is observed even if the degree of synapse loss is substantial ($>80\%$).
\par
We conclude that synapse loss, as observed in Alzheimer's disease, tends to reduce the perturbation sensitivity of the affected networks, and may thereby impair their computational performance for a broad range of task modalities.
Homeostatic mechanisms that preserve the average network activity (firing rate) can prevent this reduction in sensitivity and, hence, the decline in computational capability.

\par
So far, the reported results on the perturbation sensitivity were obtained by network simulations for a specific set of parameters.
In the following, we employ an analytical approach; firstly, to show that our findings are general and do not depend on the details of the network model, and secondly, to shed light on the mechanisms underlying the reduction in perturbation sensitivity by synapse loss and its maintenance by firing rate homeostasis.
\par
As shown in \citet{Sompolinsky88_259}, the dynamics of large random networks of analog nonlinear neurons without (or with constant) external input undergoes a transition from a stable to a chaotic regime at some critical synaptic coupling strength.
The study further revealed that this transition coincides with a critical point where the local linearized network dynamics becomes unstable. 
For more realistic networks of spiking neurons, networks with fluctuating external input or networks with a more realistic connectivity structure, a strict correspondence between the onset of chaotic dynamics and linear instability could not be established  \citep{Ostojic14,engelken15_017798,ostojic15_020354,Kadmon15_041030,Harish15_e1004266,Schuecker17_arxiv_v3}.
Nevertheless, various previous studies suggest that the two transition types are interrelated, in the sense that a change in the linear stability characteristics is accompanied by a change in the network's sensitivity to small perturbations.
\par
Here, we propose that the linear stability characteristics can serve as an indirect and easily accessible indicator of the network's sensitivity to small perturbations, and hence its computational capability.
As described in \cref{subsec:Stability-analysis}, the linearized network dynamics is determined by the effective connectivity matrix $\mat{W}$. 
Its components $w_{ij}=\mat{W}_{ij}$ ($i,j\in\{1,\ldots,N\}$) measure the effect of a small fluctuation in the firing rate $\nu_j(t)$ of a presynaptic neuron $j$ on the rate $\nu_i(t)$ of the postsynaptic neuron $i$ at a specific working point determined by the stationary firing rates $\vec{\nu}=(\nu_1,\ldots,\nu_N)$.
The effective connection weights are hence determined not only by the synaptic weights $J_{ij}$, but also by the excitability of the target cell $i$, which is in turn determined by the statistics of the synaptic input fluctuations, i.e.~the dynamical state of the local network.
The linearized dynamics becomes unstable if the spectral radius $\rho=\Real(\lambda_\text{max})$, the real part of the maximal eigenvalue $\lambda_\text{max}$ of $\mat{W}$, exceeds unity.
\par
Loss of $\EE$ synapses corresponds to setting a fraction of the excitatory components $w_{ij}$ ($i,j\in\Epop$) to zero.
In the absence of homeostatic compensation, we expect this weakening of positive feedback to have a stabilizing effect.
The dependence of the effective weights $w_{ij}$ on the working point, however, leads to a non-trivial effect of synapse loss and firing rate homeostasis on the spectral radius $\rho$.
Here, we compute $\rho$ by employing the diffusion approximation of the leaky integrate-and-fire neuron and random-matrix theory (for details, see \cref{subsec:Stability-analysis}).
\par
As shown in \cref{fig:Sensitivity-to-perturbation} (black and gray curves), the linear stability characteristics (as measured by the spectral radius $\rho$) bear striking similarities to the sensitivity to perturbations.
In the absence of homeostasis, loss of $\EE$ synapses leads to a fast decrease in $\rho$. 
Linearly unstable networks quickly become stable (\cref{fig:Sensitivity-to-perturbation}\,A).
Firing rate homeostasis, in contrast, preserves the spectral radius $\rho$, even if a substantial fraction of $\EE$ synapses is removed.
Linearly unstable networks remain unstable (\cref{fig:Sensitivity-to-perturbation}\,B), until the homeostatic resources are exhausted (\cref{fig:Sensitivity-to-perturbation}\,C).
\par
\begin{figure}[ht!]
  \centering
  \includegraphics[width=0.95\textwidth]{\figdir/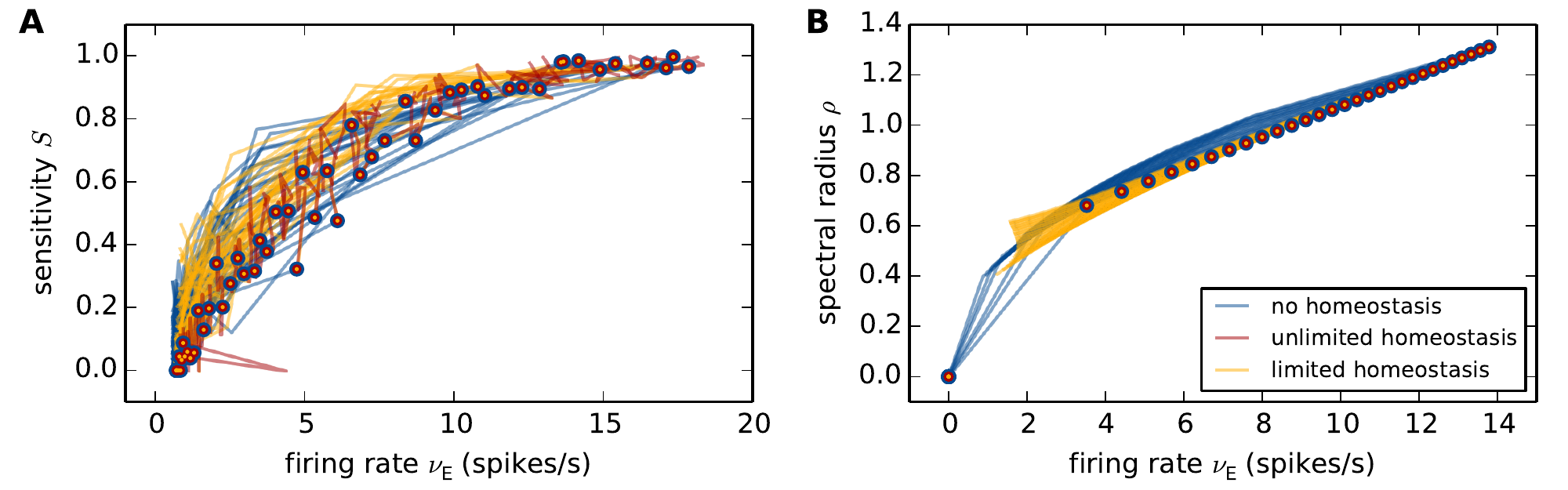}
  \caption{%
    \textbf{Firing rate as predictor of perturbation sensitivity and linear stability.}
    Dependence of the perturbation sensitivity $S$ (\textbf{A}; simulation results) and the linear stability quantified by the spectral radius $\rho$ (\textbf{B}; theory) on the mean stationary firing rate $\nuE$ of the excitatory neuron population in the absence of homeostasis (blue), as well as for unlimited (red) and limited firing rate homeostasis (yellow).
    Each curve depicts data for a fixed reference weight $J\in\{0,\ldots,3\}\,\mV$ and various degrees of synapse loss from $0\%$ to $50\%$ (vertical paths in \cref{fig:Sensitivity-to-perturbation} and \cref{fig:theory}).
    For unlimited homeostasis, the firing rate and the spectral radius are nearly perfectly conserved while removing synapses. In B), the red curves are therefore too short to be visible.
    Circles depict results for $0\%$ synapse loss.
    Same data as in \cref{fig:Sensitivity-to-perturbation} and \cref{fig:theory}\,D--F and M--O.
  }
  \label{fig:sensitivity_vs_rate_sim_and_theory}
\end{figure}

The analytical approach described in \cref{subsec:Stability-analysis} provides us with an intuitive understanding of why and under what conditions firing rate homeostasis preserves the linear stability characteristics in the face of synapse loss.
The analysis shows that, in the presence of firing-rate homeostasis, the spectral radius $\rho$ is uniquely determined by the stationary average firing rate (red curves and symbols in \cref{fig:sensitivity_vs_rate_sim_and_theory}B and eq. \cref{eq:spectral_radius_c}).
For the parameters chosen in this study, an approximately unique dependence on the firing rate is also observed in the absence of homeostasis and for limited homeostasis (blue and yellow curves in \cref{fig:sensitivity_vs_rate_sim_and_theory}B).
Network simulations reveal similar findings for the perturbation sensitivity $S$ (\cref{fig:sensitivity_vs_rate_sim_and_theory}A).
For unlimited homeostasis, the firing rate, the perturbation sensitivity and the spectral radius remain (approximately) constant during synapse loss (red curves in \cref{fig:sensitivity_vs_rate_sim_and_theory}).
In the absence of homeostasis or for limited homeostasis, firing rates change; the corresponding spectral radii $\rho$ and perturbation sensitivities $S$ nevertheless remain within a narrow band (bue and yellow curves in \cref{fig:sensitivity_vs_rate_sim_and_theory}).
The number $\KEE$ and the strength $\JEE$ of $\EE$ synapses therefore play only an indirect role by determining the stationary firing rate $\nu$.
Any combination of $\KEE$ and $\JEE$ that preserves $\nu$ will simultaneously preserve $\rho$ (and $S$).
\par
The unique dependence of the spectral radius $\rho$ on the firing rate $\nu$ is a consequence of the working-point dependence of the effective weights $w_{ij}\approx\eta(\nu_i)\J_{ij}/\sigma_i$,
where $\eta(\nu_i)$ is a function of the firing rate $\nu_i$ of the target neuron $i$ (see eq.~\cref{eq:eta}).
To maintain the stationary firing rate $\nuE$ of excitatory neurons, the synaptic weights $\JEE$ are increased to compensate for the loss of excitatory synapses, i.e.~for the decrease in the number $\KEE$ of excitatory inputs.
This increase in the synaptic weights $\J_{ij}$ (for neurons $i,j$ both in the excitatory population) is accompanied by an increase in the variance $\sigma^{2}_i$ of the synaptic input received by the target neuron $i$.
If the response firing rate $\nu_i$ is kept constant (as is the case in the presence of firing rate homeostasis), an increase in $\sigma_i$ leads to a decrease in neuron $i$'s sensitivity to a modulation of the input current caused by a spike of the source neuron $j$. This interplay between an upscaling of the weights $\J_{ij}$ and a downscaling of the neuron's modulation sensitivity restricts the growth in the effective weight $w_{ij}$, and, ultimately, leads to a preservation of the spectral radius $\rho$.
In \cref{subsec:Stability-analysis}, we demonstrate this effect for a homogeneous network of leaky-integrate-and-fire neurons.
The derivation relies on the assumption that the synaptic weights are sufficiently small and the rate of synaptic events is high (diffusion approximation), that the stationary firing rates $\nuE$ and $\nuI$ of excitatory and inhibitory neurons are identical (homogeneity), and that the input fluctuations caused by external sources are small compared to those generated by the local network.


\section{Discussion}
\label{sec-discussion}

In this article, we study the effect of Alzheimer's disease on the dynamics and perturbation sensitivity of recurrent neuronal networks.
To this end, we employ a computational model of a generic neuronal network composed of excitatory and inhibitory spiking neurons.
Alzheimer's disease is implemented in the form of a loss of excitatory synapses on to excitatory neurons.
The resulting decrease in the firing rate is avoided (or retarded) by firing rate homeostasis, which is achieved by increasing the weights of the remaining excitatory-excitatory ($\EE$) synapses. In one scenario, we allow synaptic weights to grow without bounds; in another, to ensure that they stay within the physiological range \citep{Scheff2006}, we limit the maximum synaptic weight during homeostasis to 120\% of the reference weight in the intact network (i.e. before synapse loss).

We show that, in the absence of homeostatic compensation, a progressive loss of $\EE$ synapses not only reduces the average firing rate, but also leads to an increase in spike train regularity and a decrease in the fluctuations of the population activity. At first glance, the reduction in firing rate appears to be in contrast with the observation that the network activity of early affected areas (e.g.~hippocampus) is enhanced \citep{Mendez1994,Amatniek2006,Lam2017}. However, this hyperactivity is mostly reported in very early preclinical disease stages, in which increasing oligomeric Amyloid-beta (A\text{$\beta$}) accumulates \citep{Hall2015}.  A\text{$\beta$}  oligomers seem to enhance 
the occurence of phosphorylated tau in spines \citep[for review see ][]{Tampellini2015}, causing a degradation of 
excitatory-excitatory connections \citep{Merino-Serrais2013}.  This might be the 
reason why during later disease stages, in which the tau-pathology becomes more prominent, the network activity decreases as predicted by our model \citep[see, e.g.,][]{Dickerson2005,OBrien2010,Herholz2010,Busche2012}. 

According to our AD model, the decrease in firing rate in more advanced disease stages can be delayed by homeostatic synaptic scaling. Moreover, our model predicts that as long as the homeostatic mechanisms are able to restore the network's firing rate, the CV and Fano factor are also preserved. Once these mechanisms are exhausted in the later disease stages, our model predicts that the spike train regularity increases and the fluctuations in the population activity decrease. That a weakening  of synaptic coupling decreases the CV has also been found in other computational studies \citep{Ostojic14,Kriener2014_136}; an experimental investigation of the evolution of activity statistics in AD animal models has, to our knowledge, yet to be performed.

In addition to the effects on the activity statistics, we demonstrate that the loss of synapses results in a reduction of the network's sensitivity to small perturbations, which goes hand in hand with an increase in linear stability. 
In the presence of unlimited firing rate homeostasis, the perturbation sensitivity, as well as all other dynamical network characteristics are preserved, even if the extent of synapse loss is substantial. In addition to the dynamical features, the total synaptic contact area, which is decreased in the AD network due to
synapse loss, is largely retained. 
If the homeostatic synapse growth is limited, the network dynamics as well as the total synaptic area are preserved as long as the firing rate can be maintained.
Beyond this point, the network quickly approaches the state of the pathological AD network without homeostasis. 
The effectiveness of homeostatic compensation investigated  in this study provides a possible explanation for why morphological disease-related changes in the brain (e.g. synapse loss) precede any clinically recognizable cognitive deficits by years or even decades \citep{Morris2005}.
That homeostasis is able to recover all network characteristics is non-trivial, because in the homeostatic network with few but strong $\EE$ synapses, the statistics of the synaptic input (mean and variance) is altered with respect to the intact reference network with many weak $\EE$ synapses.

\par

In order to investigate this observation further, we analyze the  linear stability characteristics  of the network and find a unique dependency of the network's spectral radius on the network's firing rate under unlimited homeostasis. Previous theoretical studies have shown that simple recurrent neuronal networks exhibit optimal computational performance for a variety of tasks if they operate in a regime where small perturbations are neither amplified nor instantly forgotten, i.e. close to the edge of chaos \citep{Langton90_12,Legenstein07_127,Legenstein07_323,Schrauwen09_1425,Dambre12_514,Schuecker17_arxiv_v3}. 
Here, we regard the network's sensitivity to a small perturbation as an indicator of its computational performance in a broad sense.
Assuming that a healthy network acts close to the edge of chaos, our results suggest that the EE-synapse loss observed in AD moves the dynamics of the network away from that point towards a less sensitive regime with stable dynamics. 

This key prediction of our study can be tested experimentally in animal models by analyzing time series of recorded neuronal activity. The degree of chaoticity can be revealed by the application of metrics such as the power spectrum, autocorrelation function, fractal dimension, Lyapunov's exponent \citep[for review see][]{Golovko2002,Beggs03_11167,Kriener2014_136}, and the analysis of neuronal avalanches \citep{Friedman12,Kriener2014_136,Beggs03_11167}.

Whereas our analysis accounts for why the sensitivity to perturbation recovers
under unlimited homeostasis, it is notable that the coefficient of variation and
the Fano factor of the spike trains are also preserved, suggesting a relationship between the transition from the
stable to the chaotic regime and these two network activity characterizations.  It has previously been proposed that the transition in spiking neuronal networks from the homogeneous asynchronous state (small sensitivity to perturbation and small CV) to the heterogeneous asynchronous state (high sensitivity to perturbation and high CV) of spiking networks is equivalent to the point where analogous rate networks become chaotic \citep{Ostojic14,Wieland2015}. Such a relationship would also explain our observation that the maintenance of the stability of the linearized network dynamics coincidences with the maintenance of the CV.

Our results raise the question of why a shift towards more stable dynamics would be disadvantageous for the system. A network that is insensitive to perturbation in the input is prone to fading memory (changes in the external input are fast forgotten, see, e.g., \citealt{Boyd85,Bertschinger04_1413}). Such networks are likely to be less flexible in responding to new inputs and thus harder to train than networks with chaotic dynamics. \citep[see FORCE learning and  liquid state computing;][]{Sussillo09_544, Maass02_2531}. On the other hand, insensitivity to small perturbations makes the system less susceptible to disruption by noise and is a prerequisite for the formation of stable attractors, which have been frequently used as memory storage embodiment in neuronal networks. \citep[e.g.,][]{Li2015}. However, more recent recordings in prefrontal and association cortices revealed that single cells exhibit complex and variable dynamics with respect to stimulus representation \citep{Jun2010}, which neither supports the hypothesis of stable attractors nor points to a network dynamics in the stable regime. 
Computational studies that have investigated the memory capacity whilst taking heterogeneous neural dynamics into account have found that memory formation succeeds well in a chaotic regime \citep{Pereira2018,Barak13_214} or with an embedding of stable subspaces in chaotic dynamics \citep{Murray16_201619449}. In addition, the construction of associative memory based on unstable periodic orbits of chaotic attractors has been suggested as a possible way of increasing  memory capacity \citep{Wagner_02}.
Thus, stable attractors and dynamics are not in line with experiments and might even be disadvantageous for memory formation. 

On the cognitive level, these results suggest that, as homeostatic compensation mechanisms begin to fail, the shift of dynamics towards the stable regime would cause a decrease in performance within a variety of domains.
For example, deficits in memory, known to primarily affect recent experiences of the AD individual, could be accounted for by the hypothesis that chaotic dynamics are needed to form new attractors \citep{Barak13_214}. In addition, very stable dynamics hinder the transition from one attractor to another, which might explain the difficulties of AD patients to perform tasks switching and dual task processing \citep{Belleville2008, Baddeley2001}. Finally, the observation that AD patients often show repetitive speech and actions \citep{Cullen2005} might be explained by difficulties in moving away from the corresponding attractor state.

\par


So far, only a few other studies on this abstraction level exist that investigate the relationship of the physical symptoms of Alzheimer's disease to its cognitive deficits.  With respect to memory, the effect of synapse loss and compensation through maintaining the TSCA has been investigated in a associative memory model \citep{Ruppin1994,Horn1993,Horn1996}. In accordance with our results, the impairment of memory retrieval due to (excitatory) synapse loss was shown to be successfully compensated by restoring the TSCA, if the restoration occurs sufficiently quickly. The effect of the restoration on the firing rate was not explicitly shown. Although these studies demonstrated that homeostasis via upscaling synapses can retain memory performance, they lack a systematic investigation of different network parameters and do not provide an analytical explanation for the results.

Apart from AD-related computational studies, the computational consequences of intrinsic and synaptic scaling-based, homeostasis has been investigated in previous studies as response to changes in the external input \citep{Naude2013, Fröhlich2008}. Based on a rate network, it has been demonstrated that intrinsic homeostasis, which shifts the neurons' transfer functions, moves the network dynamics towards the chaotic regime, stabilizes network activity in the present of Hebbian synaptic plasticity and improves input separability in response to an increasing external input \citep{Naude2013}. With respect to EE-synaptic scaling, the empirical study by \cite{Fröhlich2008} showed that if the deafferentation of the external input exceeds a certain threshold, slow periodic oscillations occur, which are also observed in several CNS disorders. It is not possible to directly compare these results to our findings, since in both studies it is changes in the external input, and not the loss of recurrent connections in the network, that triggers the homeostasis response. However, both studies are in accordance with ours on the beneficial role of the homeostasis.

\par
We complement our numerical results by an analytical approach to gain an intuitive understanding of the  mechanisms underlying the recovery of the perturbation sensitivity   (and hence, computational performance) by firing rate homeostasis. To study the linear stability characteristics of the network, we apply mean-field theory, similar to the approach used by \citet{Ostojic14}.
Note that we do not claim that a loss of linear stability coincides with the transition from stable to chaotic dynamics \citep{engelken15_017798,ostojic15_020354,Kadmon15_041030,Harish15_e1004266,Schuecker17_arxiv_v3}, as observed in large autonomous random networks of analog neurons \citep{Sompolinsky88_259}.
Rather, we exploit that the linear stability characteristics follow a similar trend as the perturbation sensitivity.
Assessing the linear stability characteristics relies on the knowledge of the effective connection strengths, i.e.~the number of excess response spikes evoked by an additional input spike in the presence of synaptic background activity.
This effective connectivity can be obtained experimentally (see, e.g., \citealp{Boucsein09_1006,London10_123}), or, for a specific neuron and synapse model, numerically (see, e.g., \citealp{Nordlie10,Heiberg13_359,Heiberg18}).
For simplified models, such as the leaky integrate-and-fire neuron studied here, it can be calculated analytically under simplifying assumptions (diffusion approximation; \citealp{Fourcaud02,Schuecker15_transferfunction}). However, we note that the preservation of linear stability by firing-rate homeostasis  is due to the approximately exponential shape of the gain function. It remains to be investigated whether our results can be generalized to other types of neurons with different gain functions. 
Our theoretical analysis exposes the working-point dependence of the effective weights as the essential mechanism underlying the recovery of linear stability by firing rate homeostasis: on the one hand, the upscaling of $\EE$ synaptic weights required for maintaining the firing rates contributes to a destabilization of the network dynamics.
On the other hand, the increase in synaptic weights leads to an increase in the variance of the synaptic-input fluctuations, which, in turn, reduces the neurons' susceptibility to modulations in the presynaptic input, and therefore stabilizes network dynamics.
Note that a similar effect has been described in \citep{Grytskyy13_258}.
\par
The present study shows that certain cognitive deficits in Alzheimer's disease may be attributed to changes in the stability characteristics of neuronal network dynamics.
Its central aim is to contribute a deeper insight into the relationship between disease related alterations at the structural, the dynamical and the cognitive level.
The findings of this study are however also applicable in an entirely different context:
in the face of limited computational resources, neuronal network models are often downscaled by reducing the number of nodes or the number of connections while increasing their strength.
This downscaling has limitations if dynamical features such as the temporal structure of correlations in the neuronal activity are to be maintained \citep{Albada15}.
The present work demonstrates that certain functional characteristics such as the sensitivity to perturbations or the classification performance can be largely preserved, if the synaptic weights are not limited by biological constraints.
This insight may be particularly relevant for cognitive-computing applications based on recurrent neuronal networks implemented in neuromorphic hardware \citep{Furber16_e51001}.
Here, the realization of natural-density connectivity and communication constitute a major bottleneck, whereas the strength of connections is hardly limited.

\par
The results reported in this study are based on a model of AD where synapse loss and synaptic scaling are confined to connections between excitatory neurons (EE). 
The motivation for restricting our investigation to the loss of $\EE$ connections is that this appears to be a prominent feature in many cortical areas \citep{Lacor07_796,Merino-Serrais2013,Dorostkar2015}. Evidence that other types of synapses are also damaged in the course of the disease has been gathered from several mouse models. For example, inhibitory synapses from neurons in the entorhinal cortex to excitatory CA1 hippocampal have been found to be selectively degenerated in AD mice \citep{Yang2016}.
Our mean field theoretical results suggest that a global unspecific synapse loss affecting all types of connections (EE, EI, IE, II) leads to noticeable changes in firing rates and linear stability characteristics, but only for higher levels of synapse loss (more than 50\%; see Supplements \cref{sec:unspecific_synapse_loss_with_homeostasis}). In this scenario, a recovery of firing rates by a synapse unspecific scaling of synaptic weights largely preserves the linear stability characteristics, similar to our findings obtained for a EE-synapse loss and EE-synapse scaling.
This suggests that the commonly reported scaling of EE-synapses may well be a mechanism the brain employs to 
compensate for alterations in dynamical characteristics that are induced by other types of
synapse loss.

\par
Although synapse loss correlates best with the cognitive decline observed in AD, by focusing on this aspect, the current study neglects other physical manifestations of AD such as neuron death and alterations of intrinsic neuronal properties  \citep{Hoxha2012,Haghani2012,Liu2013,Corbett2013,Eslamizade2015}. These phenomena would affect both inhibition and excitation in the network, so the
changes of the resulting firing rate may well be non-monotonic, unlike in our model, having unpredictable effects on the
the computational properties. Alternatively, they might be entirely unaffected: in a computational study, \citet{Barrett2016} showed that under some circumstances, a network can compensate for neuron loss without the need for additional homeostasis mechanisms by adjusting neuronal transfer functions. The contribution of intrinsic neuron contributions to the claimed hyperexcitability of inhibitory neurons observed in AD has been previously investigated  in a computational study by \cite{Perez2016}. Whereas the interplay of such properties with synaptic loss and homeostasis are out of scope of the current work, our model could be extended to incorporate these aspects. However, there is as yet no consensus on which cell type shows hyperactivity \citep{Zilberter2013}
or hypoactivity \citep{Yun2006}; which moreover may vary over the course of the disease \citep{Busche2012,Orban2010}.

\par
Analogously to our focus on synaptic loss to $\EE$ connections, we also restricted our investigation of firing rate homeostasis to $\EE$-synapse growth. This is motivated by the findings that intense synaptic upscaling is observed in AD and that an increase of excitatory-excitatory connections has been reported as a main compensation mechanism that increases the firing rate in hippocampal and cortical neurons after an artificially induced decrease in activity
(e.g. by blocking sodium channels (TTX) or  glutamatergic synapses or AMPAR)
\citep{Lissin1998,Obrien98b,Turrigiano98,Watt2000,Thiagarajan2005,Ibata2008,Kim2008}.
Other mechanisms that increase the network's firing rate could also be considered, e.g.
 changes in current flow of  ions \citep[e.g.][]{Desai99,Gibson2006} or moving the spike-initiation zone
  \citep{Grubb2010}. 
In order to understand the complexity of Alzheimer's disease it is important 
to study the effects of the different observed morphological alterations caused by AD, their corresponding homeostatic responses and, crucially, how they interfere with each other.  

\par
The findings of our study suggest that homeostatic synaptic scaling might be an attractive target for drug development. However, some caution is required.
Firstly, as discussed above, during early AD the neuronal activity seems to be increased, followed by a decrease. Thus, enhancing EE-synaptic scaling at the very beginning of AD manifestation could even accelerate the progression of the disease. In the later stages of the disease, supporting synaptic scaling might be beneficial, stabilizing the cognitive performance.  
Within this context, there are a variety of molecular substrates that regulate synaptic scaling, and which show altered expression patterns in AD, that could be considered as treatment targets, for example MSK1, PSD-95,  BDNF, Arc, Calcineurin, CaMK4 and Cdk5 (for reviews see \citealt{Jang2016}).
A major challenge is to determine whether the altered concentrations of these substrates are a consequence of direct AD pathology, or arise as an attempt of the organism to counteract pathology, or even a mixture of both. Thus, in addition to more comprehensive modelling investigations, further research on the exact time line of morphological changes and their functional implications is needed to identify promising therapeutic targets.


\section{Methods}
\subsection{Network model}
\label{subsec:network_modelling}

The network consists of $N=\NE+\NI$ identical leaky integrate-and-fire neurons, subdivided into a population of $\NE=1000$ excitatory and a population of $\NI=\NE/4$ inhibitory neurons.
In the intact reference network, each excitatory (inhibitory) neuron receives local
excitatory inputs from $\KEE=\epsilon\NE$ ($\KIE=\epsilon\NE$) randomly selected excitatory neurons, and
inhibitory inputs from $\KEI=\epsilon\NI$ ($\KII=\epsilon\NI$) randomly selected inhibitory neurons.
In addition, the neurons in the local circuit are driven by external excitatory inputs modeled as an ensemble of $p$ Poissonian spike trains with constant rate $\nuX$. 
Each of these external spike trains is sent to a subset of $\Kinput$ randomly selected (excitatory and inhibitory) neurons in the network.
Synaptic interactions are implemented in the form of stereotype exponential postsynaptic currents with a time constant $\tauS$. 
The strength  $J_{ij}$ of interaction between two neurons $j$ and $i$, the synaptic weight, is parameterized by the amplitude of the postsynaptic potential of neuron $i$ evoked by an incoming spike from neuron $j$. 
In the reference network, all excitatory connections and all inhibitory connections, respectively, have equal synaptic weights, i.e.~$\JEE=\JIE=J$ and $\JEI=\JII=-gJ$.
The greater number of excitatory inputs is compensated by a larger amplitude of inhibitory synaptic weights ($g=6$). 
\par
AD is implemented by a systematic removal of excitatory synapses to excitatory neuron ($\EE$ synapses), i.e.~by a reduction in the in-degree $\KEE$.
All other in-degrees ($\KIE$, $\KEI$, $\KII$) are preserved.
In the presence of firing-rate homeostasis, the removal of $\EE$ connections is compensated by increasing the weights $\JEE$ of the remaining $\EE$ synapses such that the time and population averaged firing rate $\nu=(NT)^{-1}\sum_{i=1}^N \int_{0}^{T} dt\, s_i(t)$ is preserved.
Here, $s_i(t)$ denotes the spike train generated by neuron $i$ (see below), and $T=1\seconds$ the simulation time.
The upscaling of the $\EE$ weights $\JEE$ is performed through bisectioning with an initial weight increment $\Delta{}\JEE=\JEE$.
The algorithm is stopped once the population averaged firing rate $\nu$ matches the rate of the corresponding intact reference network up to a precision of $0.5\%$.
In the case of limited homeostasis, $\JEE$ is set to $1.2J$ if the solution of the bisectioning exceeds $120\%$ of the reference weight $J$.
The weights $\JIE$, $\JEI$ and $\JII$ of all other connections are not changed by the firing-rate homeostasis.
\par
Unless stated otherwise, the network simulations are repeated for $M=10$ random realizations of network connectivity, initial conditions and external inputs for each parameter configuration.
A detailed description of the network model components, dynamics and parameters is given in the Supplementary Material (\cref{sec:suppl_network_model} and \cref{sec:suppl_parameters}). 
Simulations were performed using NEST version 2.10.0.

\subsection{Synaptic contact area and characterization of network activity}
\label{subsec:activity_metrics}

\paragraph{Relative total synaptic contact area}
We calculate the total synaptic contact area (TSCA) of the $\EE$ synapses 
as the product $\JEE\KEE$ of the $\EE$ weight $\JEE$ and the  $\EE$ in-degree  $\KEE$.  
The
\begin{equation}
  \label{eq:relativeTSCA}
  \text{relative TSCA}=\frac{\KEE\JEE}{\KEE^\text{ref}\JEE^\text{ref}}  
\end{equation}
is given by the ratio of the TSCA of the neurodegenerated network (reduced in-degree $K_\EE$) and the TSCA of the corresponding intact reference network (full in-degree $K_\EE$) with identical weights $\JIE$, $\JEI$ and $\JII$.

\paragraph{Spiking activity}
We represent the spike train $s_i(t)=\sum_k \delta(t-t_{i,k})$ of neuron $i$ ($i\in[1,N]$) as the superposition of Dirac-delta functions centered about the spike times $t_{i,k}$ ($k=1,2,\ldots$).
The spike count $n_i(t;b)$ is given by the number of spikes emitted in the time interval $[t,t+b]$.
For subsequent analyses, we further compute the low-pass filtered spiking activity $x_i(t)=\left(s_i*h\right)(t)$ of neuron $i$ as the linear convolution of its spike train $s_i(t)$ with an exponential kernel $h(t)=\exp\left(-t/\taufilter\right)\Theta(t)$ with time constant $\taufilter$ and Heaviside step function $\Theta(t)$.

\paragraph{Average firing rate}
The time and population averaged firing rate $\firingrate=(N T)^{-1} \sum_{i=1}^{N} n_i(0;T)$  is given by the total number $\sum_{i=1}^N n_i(T)$ of spikes emitted in the time interval $[0,T]$, normalized by the network size $N$ and the observation time $T=10\seconds$. 

\paragraph{Fano factor}
As a global measure of spiking synchrony, we employ the Fano factor
\begin{equation}
  \FF(b)=   \frac{\Var_t(n(t;b))}{\EW[t]{n(t;b)}}
\end{equation}
of the population spike count $n(t;b)=\sum_{i=1}^N n_i(t;b)$ for a binsize $b=10\ms$.
$\EW[t]{n(t;b)}$ and $\Var_t(n(t;b)$ denote the mean and the variance of the population spike count $n(t;b)$ across time, respectively.
Here, we exploit the fact that the variance of a sum signal $n(t)$ is dominated by pairwise correlations between the individual components $n_i(t)$, if the number $N$ of components is large (see, e.g., \citealp{Harris11_509,Tetzlaff12_e1002596}). Normalization by the mean $\EW[t]{n(t;b)}$ ensures that $\text{FF}(b)$ does not trivially depend on the firing rate or the binsize $b$. For an ensemble of $N$ independent realizations of a stationary Poisson process, $\text{FF}(b)=1$, irrespective of $b$ and the firing rate. In this work, an increase in $\text{FF}$ indicates an increase in synchrony on a time scale $b$.

\paragraph{Coefficient of variation}
The degree of spiking irregularity of neuron $i$ is quantified by the coefficient of variation  $\CV_i=\SD_k(\tau_{i,k})/\EW[k]{\tau_{i,k}}$
of the inter-spike intervals $\tau_{i,k}=t_{i,k}-t_{i,k-1}$, i.e.~the ratio between the standard deviation $\SD_k(\tau_{i,k})$ and the mean $\EW[k]{\tau_{i,k}}$. 
For a stationary Poisson point process, $\CV_i=1$, irrespective of its firing rate. CV's larger (smaller) than $1$ correspond to spike trains that are more (less) regular than a stationary Poisson process.
We measure  $\CV_i$ over a time interval $T=10\seconds$, and report the population average $\CV=N^{-1}\sum_{i=1}^N \CV_i$.

\paragraph{Sensitivity to perturbation}
We examine the sensitivity of a network to a small perturbation in the input spikes by performing two simulations with identical initial conditions and identical realizations of external inputs. In the second run, we apply a small perturbation by delaying one spike in one external Poisson input at time $t^*=400\ms$ by $\dtpert=0.5\ms$. 
As a measure of the network's perturbation sensitivity, we compute the Pearson correlation coefficient
\begin{equation}
  \label{eq:sensitivity_corrcoeff}
  R(t)=\frac{\EW[i]{\delta x_i(t) \delta{}x_i^*(t)}}{\sqrt{\EW[i]{\delta{}x_i(t)^2}\EW[i]{\delta{}x_i^*(t)^2}}}
\end{equation}
of the low-pass filtered spike responses $x_i(t)$ and $x_i^*(t)$ in the unperturbed and perturbed simulation, respectively, for each time point $t$.
Here, $\delta{}x_i(t)=x_i(t)-\EW[i]{x_i(t)}$ denotes the deviation of the low-pass filtered spike response $x_i(t)$ of neuron $i$ from the population average $\EW[i]{x_i(t)}$.
$\EW[i]{\ldots}=N^{-1}\sum_{i=1}^N \ldots$ represents the population average.
We define the time-dependent and the long-term perturbation sensitivity as $S(t)=1-|R(t)|$ (\cref{fig:perturbation_sensitivity_measure}, bottom panels) and $S=S(\tobs=10\seconds)$ (\cref{fig:Sensitivity-to-perturbation}), respectively.
An observation of $S(\tobs)=0$ indicates that the effect of the small perturbation has vanished, i.e.~that the network has stable dynamics and is insensitive to the perturbation. An observation of $S(\tobs)=1$, in contrast, corresponds to diverging spike patterns in response to the perturbation and thus chaotic dynamics.
In dynamical-systems theory and related applications, state differences are typically expressed in terms of the Euclidean distance $D=\sqrt{\sum_{i=1}^N \left[x_i(t)-x^*_i(t)\right]^2}$.
Here, we employ the (normalized) correlation coefficient $R$ instead to avoid (trivial) firing rate dependencies. Note that $D$ and $R$ are redundant in the sense that both can be expressed in terms of the moments $\EW[i]{x_i(t)x^*_i(t)}$, $\EW[i]{x_i(t)^2}$, $\EW[i]{x_i^*(t)^2}$, $\EW[i]{x_i}$ and $\EW[i]{x_i^*}$.

\subsection{Linearized network dynamics and stability analysis}
\label{subsec:Stability-analysis}

In the following, we describe the analytical approach to investigate the effect of synapse loss and firing rate homeostasis on the network's linear-stability characteristics.
To this end, we employ results obtained from the diffusion approximation of the leaky-integrate-and-fire (LIF) neuron with exponential postsynaptic currents under the assumption that the synaptic time constant $\tauS$ is small compared to the membrane time constant $\tauM$, and that the network activity is sufficiently asynchronous and irregular (mean-field theory; \citealp{Fourcaud02,Helias13_023002,Schuecker15_transferfunction}).
All parameters that are not explicitly mentioned here can be found in \cref{tab:parameters_1}.

\begin{figure}[ht!]
  \centering
  \includegraphics[width=\textwidth]{\figdir/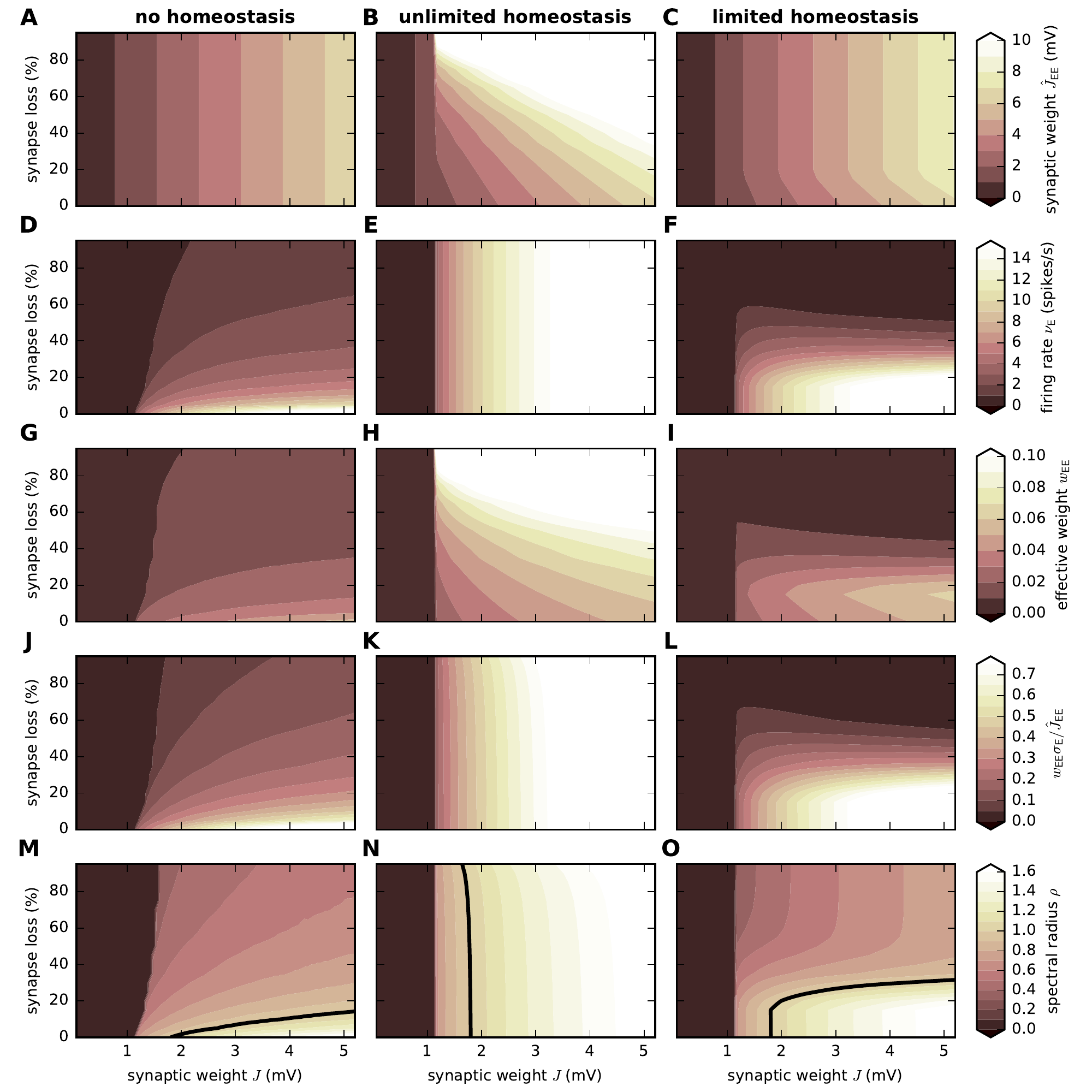}
  \caption{
    \textbf{Mean-field theory.}
    Dependence of
    the synaptic weight $\JfbEE$ (\textbf{A}--\textbf{C}),
    the average firing rate $\nuE$ of the excitatory population (\textbf{D}--\textbf{F}),
    the effective weight $\wEE$ of $\EE$ connections (\textbf{G}--\textbf{I}), 
    the ratio $\wEE\sigmaE/\JfbEE$ (\textbf{J}--\textbf{L}), 
    and the spectral radius $\rho$ (\textbf{M}--\textbf{O}) 
    on the synaptic weight $J$ and the degree of synapse loss in the absence of homeostatic compensation (left column), as well as with unlimited (middle column) and limited firing rate homeostasis (right column).
    Superimposed black curves in (M--O) mark instability lines $\rho=1$.
    Same parameters as in network simulations (see \cref{sec:suppl_network_model} and \cref{sec:suppl_parameters}).
  }
  \label{fig:theory}
\end{figure}

\paragraph{Stationary firing rates and fixed points}
For each parameter set (synaptic weight $J$, extent of synapse loss, different types of firing rate homeostasis), we first identify the self-consistent stationary states by solving
\begin{equation}
  \label{eq:selfconsistent_rates}
  \begin{aligned}
    \nuE & = G\left(\muE(\nuE,\nuI),\sigmaE(\nuE,\nuI)\right) \\
    \nuI & = G\left(\muI(\nuE,\nuI),\sigmaI(\nuE,\nuI)\right) \\
  \end{aligned}
\end{equation}
for the population averaged firing rates $\nuE$ and $\nuI$ of the excitatory and inhibitory subpopulations.
Here,
\begin{equation}
  \label{eq:siegert}
  G(\mu,\sigma)=\left( \tauR + \tauM \sqrt{\pi} \int_{\yreset}^{\ytheta} du\,f(u)\right)^{-1}
\end{equation}
represents the stationary firing rate of the LIF neuron in response to a synaptic input current with mean $\mu$ and variance $\sigma^2$ in diffusion approximation, with 
$f(u)=e^{u^2} \left[1+\erf(u)\right]$,
$\yreset=(\Vreset-\mu)/\sigma + \frac{q}{2}\sqrt{\tauS/\tauM}$, 
$\ytheta=(\theta-\mu)/\sigma + \frac{q}{2}\sqrt{\tauS/\tauM}$ 
and $q=\sqrt{2}|\zeta(1/2)|$ (with Riemann zeta function $\zeta$; \citealp{Fourcaud02,Helias13_023002,Schuecker15_transferfunction}).
For stationary firing rates $\nuE$ and $\nuI$ of the local presynaptic neurons, the mean and the variances of the total synaptic input currents to excitatory and inhibitory neurons are given by
\begin{equation}
  \label{eq:selfconsistent_mean_variance}
  \begin{aligned}
    \muE & = \left( \KEE \JfbEE \nuE + \KEI \JfbEI \nuI + \KX \JfbX \nuX \right) \tauM,\\
    \muI & = \left( \KIE \JfbIE \nuE + \KII \JfbII \nuI + \KX \JfbX \nuX \right) \tauM,\\
    \sigmaE^2 & = \left( \KEE \JfbEE^2 \nuE + \KEI \JfbEI^2 \nuI + \KX \JfbX^2 \nuX \right) \tauM,\\
    \sigmaI^2 & = \left( \KIE \JfbIE^2 \nuE + \KII \JfbII^2 \nuI + \KX \JfbX^2 \nuX \right) \tauM,\\   
  \end{aligned}
\end{equation}
respectively.
The coefficients $K_{pq}$ ($p,q\in\{\exc,\inh\}$) denote the number of inputs (in-degree) to neurons in population $p$ from population $q$, 
$\Jfb_{pq}=\tauS\CM^{-1}\PSCamp_{pq}$ the corresponding rescaled PSC amplitude, 
$\KX$ the number of external inputs for each neuron in the network, and $\nuX$ the firing rate of the Poissonian external sources.
Note that in our network simulations, each external source is connected to a randomly selected subset of $\Kinput$ neurons.
As a result, the number $\KX$ of external inputs each neuron in the network receives is a binomially distributed random number.
For the analytical treatment, we neglect this variability and replace $\KX$ by the average $\KX=p\Kinput/N$.
Equations \cref{eq:selfconsistent_rates} and \cref{eq:selfconsistent_mean_variance} are simultaneously solved numerically using the \verb+optimize.root()+ function (\verb+method=’hybr’+) of the \verb+scipy+ package (\url{http://www.scipy.org}).
To ensure that all solutions are found, the fixed-point search is repeated for $30$ pairs of initial rates randomly drawn from a uniform distribution between $0$ and $50\sps$.
If multiple coexisting fixed points are found, the one with the highest firing rates is chosen for the subsequent analysis.
\paragraph{Synapse loss and firing rate homeostasis}
In this work, Alzheimer's disease is modeled by removing a fraction of $\EE$ synapses, i.e.~by reducing the in-degree $\KEE$.
The self-consistent firing rates $\nuE$ and $\nuI$ after synapse removal are hence reduced (\cref{fig:theory}\,D).
In the presence of unlimited firing rate homeostasis, we adjust the weight $\JfbEE$ (of the remaining synapses) (\cref{fig:theory}\,B) until the excitatory self-consistent firing rate $\nuE^\text{ref}$ of the intact reference network (before synapse removal) is recovered (\cref{fig:theory}\,E).
To this end, we numerically find the roots of $\nuE-\nuE^\text{ref}$ by employing again \verb+scipy+'s \verb+optimize.root()+ function.
We repeat the root finding for $30$ initial weights randomly drawn from a uniform distribution between $\JfbEE^\text{ref}$ and $10\JfbEE^\text{ref}$, where $\JfbEE^\text{ref}$ denotes the original weight before synapse removal, and keep the solution where $|\nuE-\nuE^\text{ref}|$ is minimal.
For limited homeostasis, the new $\EE$ weight is chosen as the minimum of the solution $\JfbEE$ and $1.2\JfbEE^\text{ref}$ (\cref{fig:theory}\,C,F).

\begin{figure}[ht!]
  \centering
  \includegraphics[width=\textwidth]{\figdir/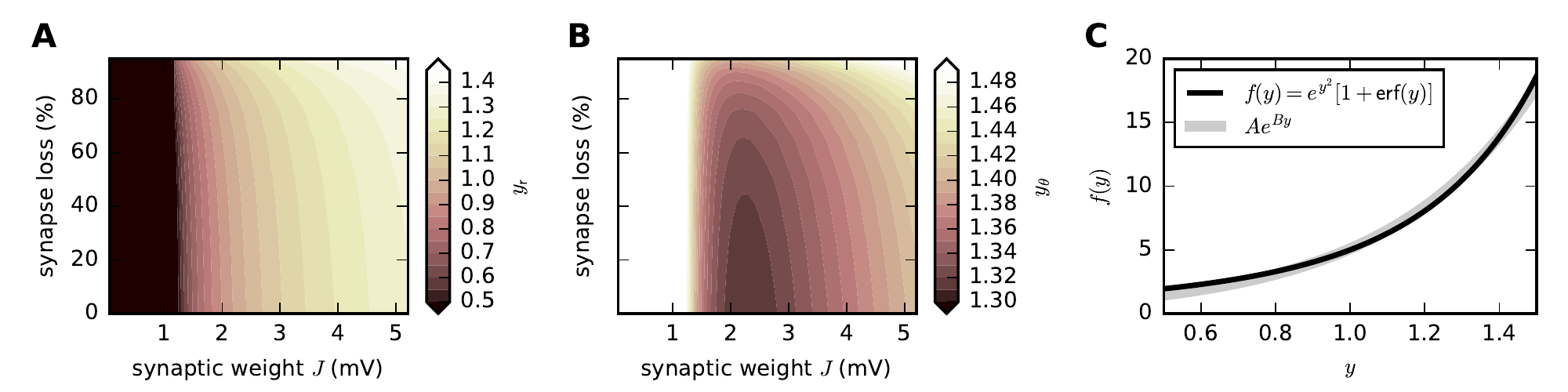}
  \caption{
    \textbf{Approximation of $f(y)=e^{y^2}[1+\erf(y)]$ by an exponential function.}
    \textbf{A},\textbf{B)} Dependence of ${\yreset}_\exc$ (A) and ${\ytheta}_\exc$ (B)
    on the synaptic reference weight $J$ and the degree of synapse loss 
    in the presence of unlimited firing rate homeostasis (mean-field theory).
    \textbf{C)} Graph of $f(y)=e^{y^2}[1+\erf(y)]$ (black) and exponential function $Ae^{BX}$ (gray; $A=0.4$, $B=2.5$) fitted to $f(y)$ in interval $y\in[0.5,1.5]$.
    Same parameters as in network simulations (see \cref{sec:suppl_network_model} and \cref{sec:suppl_parameters}).
  }
  \label{fig:theory_exponential_fit}
\end{figure}

\begin{figure}[ht!]
  \centering
  \includegraphics[width=\textwidth]{\figdir/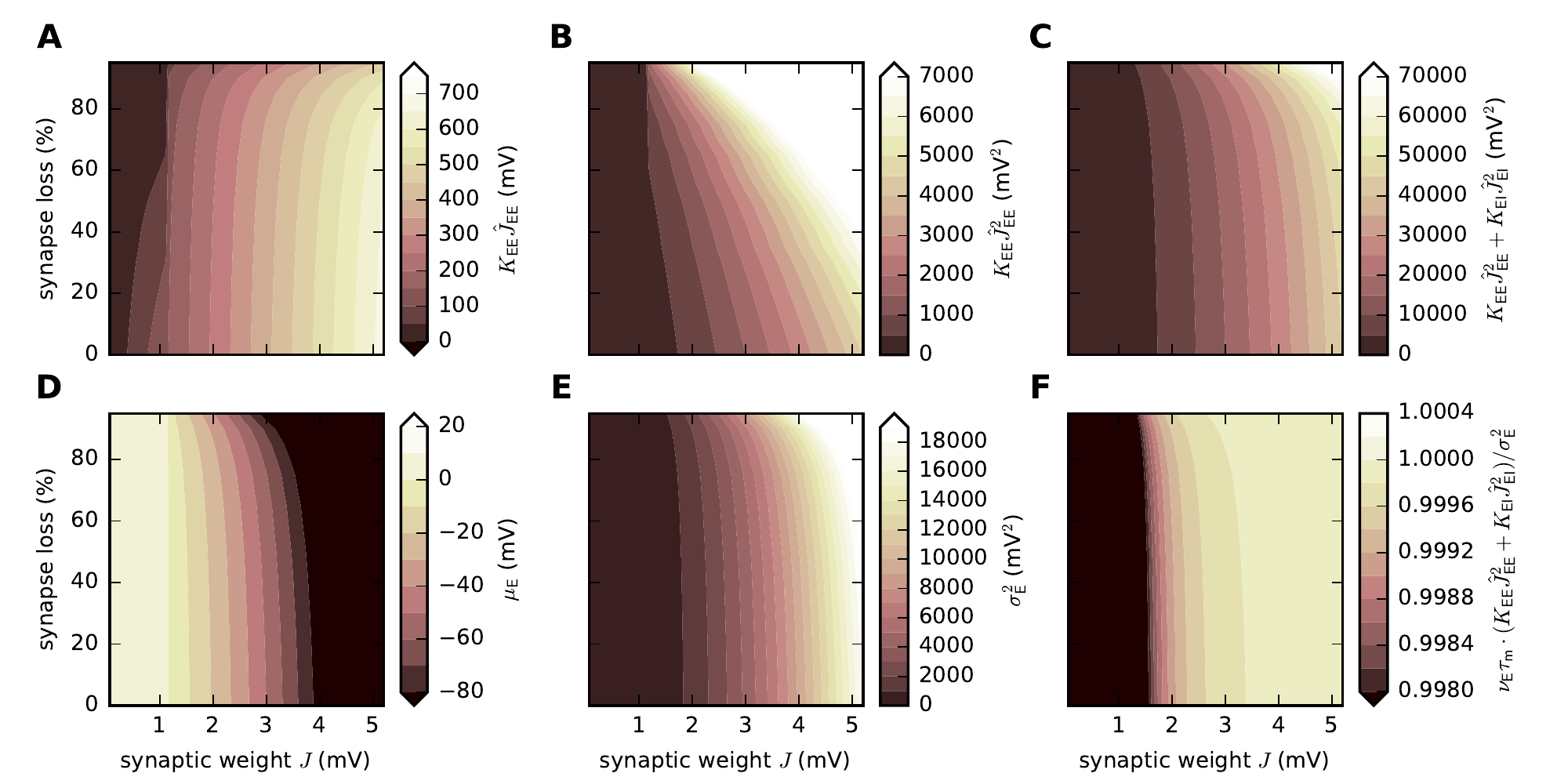}
  \caption{%
    \textbf{Canceling of the synaptic-weight variance by the input variance.}
    Dependence of 
    $\KEE\JfbEE$ (\textbf{A}),
    $\KEE\JfbEE^2$ (\textbf{B}),
    $\KEE\JfbEE^2+\KEI\JfbEI^2$ (\textbf{C}),
    input mean $\muE$  (\textbf{D}),
    input variance $\sigmaE^2$ (\textbf{E}), and
    the ratio $\nuE\tauM\,(\KEE\JfbEE^2+\KEI\JfbEI^2) / \sigmaE^{*2}$ (\textbf{F})
    on the synaptic reference weight $J$ and the degree of synapse loss 
    in the presence of unlimited firing rate homeostasis (mean-field theory).
    Note that $\nuE\tauM\,(\KEE\JfbEE^2+\KEI\JfbEI^2) / \sigmaE^{*2}$ (F) is very close to unity in all regions where $\nuE>0$ (cf.~\cref{fig:theory}\,E). Hence, the ratio between the synaptic-weight variance $\KEE\JfbEE^2+\KEI\JfbEI^2$ and the synaptic-input variance $\sigmaE^{*2}$ is uniquely determined by the firing rate.
    Same parameters as in network simulations (see \cref{sec:suppl_network_model} and \cref{sec:suppl_parameters}).
    }
  \label{fig:theory_compensation}
\end{figure}

\paragraph{Linearized network dynamics and effective connectivity}
As shown in \citep{Tetzlaff12_e1002596,Helias13_023002}, networks of spiking neurons can be formally linearized about a stationary state $\vec{\nu}^*=(\nu^*_1,\ldots,\nu^*_N)$ (linear-response theory) and thereby be mapped to an $N$-dimensional system 
\begin{equation}
  \label{eq:linearized_dynamics}
  \delta\nu_i(t) = \sum_{j=1}^N (h_{ij} * \delta\nu_j)(t)\qquad(i\in[1,N])
\end{equation}
of linear equations describing the dynamics of small firing rate fluctuations $\delta\nu_i(t)=\nu_i(t)-\nu_i^*$ around this stationary state.
The stationary states are determined as the self-consistent solutions of
\begin{equation}  
  \vec{\nu}^*=\vec{\phi}(\vec{\nu}^*),
\end{equation}
where $\vec{\phi}(\vec{\nu}_\text{in})$ represents the activation function mapping the vector of stationary input rates $\vec{\nu}_\text{in}$ to the vector of output rates.
The coupling kernel $h_{ij}(t)$ represents the firing rate impulse response, i.e.~the modulation in the output rate $\nu_i(t)$ in response to a delta-shaped fluctuation in the rate $\nu_j(t)$ of presynaptic neuron $j$.
We refer to the area 
\begin{equation}
\label{eq:integral_kernel}
  w_{ij}=\int_{-\infty}^\infty dt\,h_{ij}(t)
\end{equation}
under the coupling kernel as the effective connection weight.
It measures the average number of extra spikes emitted by target neuron $i$ in response to a spike fired by the presynaptic neuron $j$, in the context of the background activity determined by the stationary state $\vec{\nu}^*$.
Exploiting the fact that the integral of the impulse response of a linear(ized) system is identical to the long-term limit of its step response, the effective weight
\begin{equation}
  \label{eq:effective_weights}
  w_{ij}=\left.\frac{\partial\phi_i(\vec{\nu})}{\partial\nu_j}\right|_{\vec{\nu}^*}
\end{equation}
is given by the derivative of the activation function $\phi_i$ of neuron $i$ with respect to the stationary firing rate $\nu_j$ of neuron $j$, evaluated at the stationary state $\vec{\nu}^*$.
With $\phi_i(\vec{\nu})=G(\mu_i(\vec{\nu}),\sigma_i(\vec{\nu}))$ from \cref{eq:siegert}, $\mu_i(\vec{\nu})=\left(\sum_{j=1}^N \Jfb_{ij} \nu_j + \KX\JfbX\nuX\right)\tauM$, and $\sigma^2_i(\vec{\nu})=\left(\sum_{j=1}^N \Jfb^2_{ij} \nu_j + \KX\JfbX^2\nuX\right)\tauM$, we obtain
\begin{equation}
  \label{eq:effective_weights_lif}
  w_{ij}
  =\left.\frac{\partial{}G}{\partial\mu_i}\frac{\partial\mu_i}{\partial\nu_j}\right|_{\vec{\nu}^*} + \left.\frac{\partial{}G}{\partial\sigma_i}\frac{\partial\sigma_i}{\partial\nu_j}\right|_{\vec{\nu}^*}  
  =\frac{\Jfb_{ij}}{\sigma_i^*} \sqrt{\pi}(\tauM\nu_i^*)^2\left( f({\ytheta^*}_{\!i})-f({\yreset^*}_{\!\!i})\right)
\end{equation}
as the effective weight of the LIF neuron in the stationary self-consistent state given by $\vec{\nu}^*$ \citep{Tetzlaff12_e1002596,Helias13_023002}.
Note that for the result on the right-hand side of \cref{eq:effective_weights_lif}, we account only for the derivative $\frac{\partial{}G}{\partial\mu_i}$ of $G$ with respect to the mean input $\mu_i$ (DC susceptibility), but neglect the contribution $\frac{\partial{}G}{\partial\sigma_i}$ resulting from a modulation in the input variance $\sigma_i^2$.
Removal of $\EE$ synapses and the resulting decrease in stationary firing rates (\cref{fig:theory}\,D) leads to a reduction in the effective weight $\wEE$ of $\EE$ connections (\cref{fig:theory}\,G).
In the presence of (unlimited) firing rate homeostasis, upscaling of $\EE$ synapses (\cref{fig:theory}\,B) and the resulting preservation of firing rates (\cref{fig:theory}\,E) results in an increase in $\wEE$ (\cref{fig:theory}\,H).
\paragraph{Stability analysis}
For the LIF neuron with weak exponential synapses \citep{Helias13_023002} as well as for a variety of other neuron and synapse models \citep{Nordlie10,Heiberg13_359,Heiberg18},
the effective coupling kernel $h_{ij}(t)$ introduced in \cref{eq:linearized_dynamics} can be well approximated by an exponential function
$h_{ij}(t)=w_{ij}\tau^{-1}\exp(-t/\tau)\Theta(t)$ with an effective time constant $\tau$ and Heaviside function $\Theta(t)$.
With this approximation, \cref{eq:linearized_dynamics} can be written in form of an $N$-dimensional system of differential equations
\begin{equation}
  \label{eq:linearized_dynamics_ode}
  \tau\frac{d\delta\vec{\nu}}{dt} = -\delta\vec{\nu} + \mat{W} \delta\vec{\nu}(t).
\end{equation}
Here, $\mat{W}=\{w_{ij}\}$ denotes the $N\times{}N$ effective connectivity matrix and $\delta\vec{\nu}(t)=(\delta\nu_1(t),\ldots,\delta\nu_N(t))$ the vector of firing rate fluctuations.
The system \cref{eq:linearized_dynamics_ode} has bounded solutions only if the real parts of all Eigenvalues $\lambda_k$ of the effective connectivity matrix $\mat{W}$ are smaller than unity, i.e.~if $\Real(\lambda_k)<1$ ($\forall{}k$).
If $\rho=\max_{k} \left( \Real(\lambda_k)\right)>1$, the linearized system is unstable and fluctuations diverge.
In the original nonlinear LIF network, an unbounded growth of fluctuations is prevented by the nonlinearities of the single-neuron dynamics.
For large random networks where the statistics of the coupling strengths does not depend on the target nodes,
the bulk of Eigenvalues $\{\lambda_k|k\in[1,N]\}$ of $\mat{W}$ is located in the complex plane within a circle centered at the coordinate origin and a radius $\rho$ which is determined by the variances of the effective connectivity \citep{Rajan06}.
A single outlier is given by the Eigenvalue $\lambda_{k^*}$ associated with the Eigenvector $\vec{u}_{k^*}=(1,1,\ldots,1,1)\transp$, which is given by the mean effective weight.
In inhibition dominated networks, the mean synaptic weight and, hence, $\lambda_{k^*}$ are negative.
The stability behaviour is therefore solely determined by the spectral radius $\rho$.
For a random network composed of $\NE$ excitatory ($j\in\Epop$; $\NE=|\Epop|$) and $\NI$ inhibitory neurons ($j\in\Ipop$; $\NI=|\Ipop|$) with homogeneous in-degrees $K_{pq}$ ($p,q\in\{\exc,\inh\}$) and weights
\begin{equation}
  w_{ij}=
  \begin{cases}
    \wEE & \forall i\in\Epop,\,j\in\Epop,\ \text{connection $j\to{}i$ exists with probability}\ \frac{\KEE\NE}{N\NE}\\
    \wEI & \forall i\in\Epop,\,j\in\Ipop,\ \text{connection $j\to{}i$ exists with probability}\ \frac{\KEI\NE}{N\NI}\\
    \wIE & \forall i\in\Ipop,\,j\in\Epop,\ \text{connection $j\to{}i$ exists with probability}\ \frac{\KIE\NI}{N\NE}\\
    \wII & \forall i\in\Ipop,\,j\in\Ipop,\ \text{connection $j\to{}i$ exists with probability}\ \frac{\KII\NI}{N\NI}\\
    0    & \forall i,j,\ \text{connection $j\to{}i$ does not exist}
  \end{cases},
\end{equation}
the squared spectral radius is given by
\begin{equation}
  \label{eq:spectral_radius}
    \rho^2
    =\NE v_\exc + \NI v_\inh
    =N^{-1}\left(\KEE\NE\wEE^2 + \KIE\NI\wIE^2 + \KEI\NE\wEI^2 + \KII\NI\wII^2 \right).    
\end{equation}
Here,  
$v_\exc=\wEE^2\KEE\NE/(N\NE) + \wIE^2\KIE\NI/(N\NI)$ 
and 
$v_\inh=\wEI^2\KEI\NE/(N\NI) + \wII^2\KII\NI/(N\NI)$ 
denote the variances of the effective connectivity $w_{ij}$ across the ensemble of target cells ($i\in[1,N]$) for excitatory ($j\in\Epop$) and inhibitory sources ($j\in\Ipop$), respectively.
Without homeostatic compensation, $\EE$-synapse loss leads to a stabilization of the linearized network dynamics, i.e.~a decrease in $\rho$ (\cref{fig:theory}\,M).
In the presence of unlimited firing rate homeostasis, the spectral radius $\rho$ is preserved (\cref{fig:theory}\,N), even if a substantial fraction of $\EE$ synapses is removed  (\cref{fig:theory}\,N).
If the homeostatic resources are limited, $\rho$ is maintained until the upscaled synaptic weights reach their maximum value (\cref{fig:theory}\,O).
\paragraph{Preservation of linear stability by firing rate homeostasis}
At first glance, it is unclear why firing rate homeostasis preserves the linear stability characteristics as measured by the spectral radius $\rho$.
While the stationary firing rates $\nu_i^*$ are, by definition, kept constant during synapse loss and homeostasis, the input statistics $\mu_i^*$, $\sigma_i^*$ (\cref{fig:theory_compensation}\,D,E), ${\yreset^*}_{\!\!i}$ and ${\ytheta^*}_{\!i}$ (\cref{fig:theory_exponential_fit}\,A,B) as well as the effective weights $\effweight_{ij}$ (\cref{fig:theory}\,K) are not.
To shed light on the mechanisms leading to the preservation of $\rho$, we first note that the factor
\begin{equation}
  \label{eq:eta}
  \sqrt{\pi}(\tauM\nu^*_i)^2\left( f({\ytheta^*}_{\!i})-f({\yreset^*}_{\!\!i})\right)\eqqcolon\eta(\nu_i^*)
\end{equation}
on the right-hand side of \cref{eq:effective_weights_lif} is in good approximation uniquely determined by the stationary firing rate $\nu_i^*$ (\cref{fig:theory}\,J--L).
This can be understood by noting 
that, according to \cref{eq:siegert}, the firing rates are determined by 
\begin{math}
  \int_{{\yreset^*}_{\!\!i}}^{{\ytheta^*}_{\!i}} dy\, f(y),  
\end{math}
and that $f(y)=e^{y^2}[1+\erf(y)]$ can be approximated by an exponential function $f(y)\approx{}Ae^{By}$ for the range of arguments spanned by ${\yreset^*}_{\!\!i}$ and ${\ytheta^*}_{\!i}$ (\cref{fig:theory_exponential_fit}). 
With this approximation, 
\begin{math}
   \int_{{\yreset^*}_{\!\!i}}^{{\ytheta^*}_{\!i}} dy\, f(y) = B^{-1} \left[f({\ytheta^*}_{\!i}) - f({\yreset^*}_{\!\!i})\right]
\end{math}.
For constant firing rate, $f({\ytheta^*}_{\!i}) - f({\yreset^*}_{\!\!i})$ is therefore constant, too, and 
the effective weight is essentially determined by the ratio $\Jfb_{ij}/\sigma_i^*$.
With $w_{pq}=\eta(\nu^*_p)\Jfb_{pq}/\sigma^*_p$ ($p,q\in\{\exc,\inh\}$), \cref{eq:spectral_radius} reads
\begin{equation}
  \label{eq:spectral_radius_b}
  \begin{aligned}
  \rho^2
  &=N^{-1}\left(
    \KEE\NE\frac{\JfbEE^2}{\sigmaE^{*2}}\eta^2(\nuE^*) + 
    \KIE\NI\frac{\JfbIE^2}{\sigmaI^{*2}}\eta^2(\nuI^*) + 
    \KEI\NE\frac{\JfbEI^2}{\sigmaE^{*2}}\eta^2(\nuE^*) + 
    \KII\NI\frac{\JfbII^2}{\sigmaI^{*2}}\eta^2(\nuI^*) 
  \right)\\
  &=N^{-1}\left(
    \eta^2(\nuE^*)\NE\frac{\KEE\JfbEE^2+\KEI\JfbEI^2}{\sigmaE^{*2}} +
    \eta^2(\nuI^*)\NI\frac{\KIE\JfbIE^2+\KII\JfbII^2}{\sigmaI^{*2}}
  \right).
  \end{aligned}
\end{equation}
According to our network simulations as well as the mean-field theory described above, stationary firing rates of the excitatory and inhibitory subpopulation are identical in the presence of firing rate homeostasis, i.e.~$\nu^*\coloneqq\nuE^*=\nuI^*$. With \cref{eq:selfconsistent_mean_variance} and assuming that the contribution $\KX\JfbX^2\nuX$ of the external drive to the total input variances $\sigma^{*2}_{\exc/\inh}$ can be neglected (which is the case for the range of parameters considered in this study), we find that the spectral radius
\begin{equation}
  \label{eq:spectral_radius_c}
  \begin{aligned}
    \rho^2
    = \frac{\eta^2(\nu^*)}{\nu^*\tauM}
  \end{aligned}
\end{equation}
is in good approximation uniquely determined by the stationary firing rate $\nu^*$ (\cref{fig:theory_compensation} and \cref{fig:sensitivity_vs_rate_sim_and_theory}B). A constant firing rate (as achieved by firing rate homeostasis) is therefore accompanied by a constant spectral radius. 

\section{Author Contributions}
The spiking neural network model and the data analyses were developed by CB, TT and AM. CB implemented the model and carried out the numerical simulations and the data analysis. The mean-field model was developed, implemented and analyzed by CB and TT. RD investigated the network with respect to computational capacity. The manuscript was written by CB, TT, RD and AM.

\section{Acknowledgments}
We thank Susanne Kunkel and Philipp Bamberger for their conceptual and technical support during the project, 
as well as Sven Goedeke, Raoul-Martin Memmesheimer, Moritz Helias, and Robert Legenstein for fruitful discussions.
This project has received funding from 
the Initiative and Networking Fund of the Helmholtz Association, 
Helmholtz Portfolio Theme ``Supercomputing and Modeling for the Human Brain''), and
the European Union's Horizon 2020 Framework Programme for Research and Innovation under Specific Grant Agreement No.~720270 (Human Brain Project SGA1 and SGA2).


\clearpage
\section{Supplementary Materials}
\label{sec:supplemental_material}
\subsection{Network model}
\label{sec:suppl_network_model}

\begin{table}[H]
\renewcommand{\arraystretch}{1.2}
\begin{tabular}{|@{\hspace*{1mm}}p{3cm}@{}|@{\hspace*{1mm}}p{12cm}|}
\hline 
\multicolumn{2}{|>{\color{white}\columncolor{black}}c|}{\textbf{Summary}}\tabularnewline
\hline 
\textbf{Populations} & excitatory population $\Epop$, inhibitory population $\Ipop$ \tabularnewline
\hline 
\textbf{Connectivity} & random convergent connections (fixed in-degrees)\tabularnewline
\hline 
\textbf{Neuron model } & leaky integrate-and-fire (LIF)
\tabularnewline
\hline 
\textbf{Synapse model } & exponentially decaying postsynaptic currents, static synaptic weights, fixed delays \tabularnewline
\hline 
\textbf{Input} & Poissonian spike trains \tabularnewline
\hline 
\end{tabular}
\begin{tabular}{|@{\hspace*{1mm}}p{3cm}@{}|@{\hspace*{1mm}}p{5.95cm}@{}|@{\hspace*{1mm}}p{5.95cm}|}
\hline 
\multicolumn{3}{|>{\color{white}\columncolor{black}}c|}{\textbf{Populations}}\tabularnewline
\hline 
\textbf{Name} & \textbf{Elements} & \textbf{Size}\tabularnewline
\hline 
$\Epop$ & LIF & $N_\exc=K/\epsilon$\tabularnewline
\hline 
$\Ipop$ & LIF & $N_\inh=\gamma N_\exc=\gamma K/\epsilon$\tabularnewline
\hline 
\end{tabular}
\begin{tabular}{|@{\hspace*{1mm}}p{1.45cm}@{}|@{\hspace*{1mm}}p{1.45cm}@{}|@{\hspace*{1mm}}p{12cm}|}
\hline 
\multicolumn{3}{|>{\color{white}\columncolor{black}}c|}{\textbf{Connectivity}}\tabularnewline
\hline 
\textbf{Source} & \textbf{Target} & \textbf{Pattern}\tabularnewline
\hline 
$\Epop$ & $\Epop$ & random convergent, in-degree $K_\EE$, delay $d$, weight $J_\EE$ \tabularnewline
\hline 
$\Epop$ & $\Ipop$ & random convergent, in-degree $K_\IE=K$, delay $d$, weight $J_\IE=J$\tabularnewline
\hline 
$\Ipop$ & $\Epop$ & random convergent, in-degree $K_\EI=\gamma K$, delay $d$, weight $J_\EI=-gJ$ \tabularnewline
\hline 
$\Ipop$ & $\Ipop$ & random convergent, in-degree $K_\II=K$, delay $d$, weight $J_\II=-gJ$ \tabularnewline
\hline 
all & all & no self-connections (``autapses''), no multiple connections (``multapses'') \tabularnewline
\hline 
\end{tabular}
\begin{tabular}{|@{\hspace*{1mm}}p{3cm}@{}|@{\hspace*{1mm}}p{12cm}|}
\hline 
\multicolumn{2}{|>{\color{white}\columncolor{black}}c|}{\textbf{Neuron }}\tabularnewline
\hline 
\textbf{Type} & leaky integrate-and-fire (LIF) model\tabularnewline
\hline 
\textbf{Description} & dynamics of membrane potential $V_{i}(t)$ ($i\in\left\{ 1,\ldots,N\right\} $)
\begin{itemize}
\item spike emission at $t_{k}^{i}$ if $V_{i}\left(t_{k}^{i}\right)\geq\theta$ 
\item subthreshold dynamics:
  $\tauM\dot{V}_{i}=-V_{i}+\RM{}I_{i}(t)$\quad $\forall{}k,\ \forall t \notin \left[t_{k}^{i},\,t_{k}^{i}+\tauR\right)$
\item reset and refractoriness: 
  $V_{i}(t)=\Vreset$ \quad $\forall{}k,\ \forall t \in \left(t_{k}^{i},\,t_{k}^{i}+\tauR\right]$  
\end{itemize}
initial membrane-potential distribution at $t=0$: random uniform between $0$ and $\theta$\tabularnewline
& exact integration with continuous spike times in discrete-time simulation \citep{Rotter99a,Morrison2007,Hanuschkin10_113}\tabularnewline
& temporal resolution $\dtsim$\tabularnewline
\hline 
\end{tabular}
\begin{tabular}{|@{\hspace*{1mm}}p{3cm}@{}|@{\hspace*{1mm}}p{12.cm}|}
\hline 
\multicolumn{2}{|>{\color{white}\columncolor{black}}c|}{\textbf{Synapse}}\tabularnewline
\hline 
\textbf{Type} & current based synapses with exponential post-synaptic currents (PSCs)\tabularnewline
\hline 
\textbf{Description} & 
$I_i(t)=\sum_{j=1}^{N} \PSCamp_{ij} (\text{PSC}*s_j)(t)$\tabularnewline
& with $\displaystyle\text{PSC}(t)=e^{-t/\tauS} \Theta(t)$ and Heaviside function $\Theta(t)=\begin{cases}1 & t \ge 0 \\ 0 & \text{else} \end{cases}$\tabularnewline
& \,$\curvearrowright$ post-synaptic potential $\displaystyle\text{PSP}_{ij}(t)=\PSCamp_{ij}\frac{\RM\tauS}{\tauS-\tauM}\left(e^{-t/\tauS}-e^{-t/\tauM}\right)\Theta(t)$\tabularnewline
& \,synaptic weight $\displaystyle\synweight_{ij}=\PSCamp_{ij}\frac{\RM\tauS}{\tauS-\tauM} \left(\left[\frac{\tauM}{\tauS}\right]^{\frac{-\tauM}{\tauM-\tauS}}-\left[\frac{\tauM}{\tauS}\right]^{\frac{-\tauS}{\tauM-\tauS}}\right)=\underset{t}{\text{max}}\left(\text{PSP}_{ij}(t)\right)$
\tabularnewline
\hline 
\end{tabular}
\begin{tabular}{|@{\hspace*{1mm}}p{3cm}@{}|@{\hspace*{1mm}}p{12.cm}|}
\hline 
\multicolumn{2}{|>{\color{white}\columncolor{black}}c|}{\textbf{Input}}\tabularnewline
\hline 
\textbf{Type} & spike trains modeled as independent realizations of a Poisson point process \tabularnewline
\hline 
\textbf{Description} & $\numstims$ independent Poisson spike trains of rate $\nu_{\ext}$,  each connected to $\Kinput$ randomly chosen (excitatory and inhibitory) network neurons  \tabularnewline
\hline 
\end{tabular}
\begin{tabular}{|@{\hspace*{1mm}}p{3cm}@{}|@{\hspace*{1mm}}p{12.cm}|}
\hline 
\multicolumn{2}{|>{\color{white}\columncolor{black}}c|}{\textbf{Realizations}}\tabularnewline
\hline 
\textbf{Description} & repetition of network simulations for $M$ random realizations of network connectivity, initial conditions, and external inputs \tabularnewline
\hline 
\end{tabular}
\caption{\label{tab:Model-description} Description of the network model according to \citep{Nordlie-2009_e1000456}.}
\end{table}

\subsection{Lists of parameters}
\label{sec:suppl_parameters}

\begin{table}[H]
\renewcommand{\arraystretch}{1.2}
\begin{tabular}{|@{\hspace*{1mm}}p{3cm}@{}|@{\hspace*{1mm}}p{4cm}@{}|@{\hspace*{1mm}}p{8.1cm}|}
\hline 
\multicolumn{3}{|>{\columncolor{gray}}c|}{\textbf{Connectivity}}\tabularnewline
\hline 
\textbf{Name} & \textbf{Value } & \textbf{Description}\tabularnewline
\hline 
$\K$ & $100$ & excitatory in-degree (number of excitatory inputs) of reference network\tabularnewline
\hline 
$\KEE$ & $5,10,20,30,\ldots,100$ & excitatory-excitatory in-degree\tabularnewline
\hline 
$\epsilon$ & $0.1$ & network density\tabularnewline
\hline 
$\gamma$ & $1/4$ & relative size of inhibitory subpopulation\tabularnewline
\hline 
\end{tabular}
\begin{tabular}{|@{\hspace*{1mm}}p{3cm}@{}|@{\hspace*{1mm}}p{4cm}@{}|@{\hspace*{1mm}}p{8.1cm}|}
\hline 
\multicolumn{3}{|>{\columncolor{gray}}c|}{\textbf{Neuron}}\tabularnewline
\hline 
\textbf{Name} & \textbf{Value } & \textbf{Description}\tabularnewline
\hline 
$\tauM$ & $20\ms$ & membrane time constant\tabularnewline
\hline 
$\tauR$ & $2\ms$ & absolute refractory period\tabularnewline
\hline 
$\CM$ & $250\pF$ & membrane capacity\tabularnewline
\hline 
$\Vreset$ & $0.0\mV$ & reset potential\tabularnewline
\hline 
$\tauS$ & $2\ms$ & time constant of post-synaptic current\tabularnewline
\hline 
$\theta$ & $15\mV$ & spike threshold\tabularnewline
\hline 
\end{tabular}
\begin{tabular}{|@{\hspace*{1mm}}p{3cm}@{}|@{\hspace*{1mm}}p{4cm}@{}|@{\hspace*{1mm}}p{8.1cm}|}
\hline 
\multicolumn{3}{|>{\columncolor{gray}}c|}{\textbf{Synapse}}\tabularnewline
\hline 
\textbf{Name} & \textbf{Value } & \textbf{Description}\tabularnewline
\hline 
$\J$ & $0.05,0.1,\ldots,4.95\mV$ & EPSP amplitude\tabularnewline
\hline 
$g$ & $6$ & relative IPSP amplitude\tabularnewline
\hline 
$d$ & $1\ms$ & spike transmission delay\tabularnewline
\hline 
\end{tabular}
\begin{tabular}{|@{\hspace*{1mm}}p{3cm}@{}|@{\hspace*{1mm}}p{4cm}@{}|@{\hspace*{1mm}}p{8.1cm}|}
\hline 
\multicolumn{3}{|>{\columncolor{gray}}c|}{\textbf{Input}}\tabularnewline
\hline 
\textbf{Name} & \textbf{Value } & \textbf{Description}\tabularnewline
\hline 
$\nu_{\ext}$ & $750\sps$ & rate of external Poisson inputs\tabularnewline
\hline 
$\J_{\ext}$ & $0.2\mV$   & PSP  amplitude evoked by external inputs\tabularnewline
\hline
$\numstims$ & 5  & number of input sources (spike trains)\tabularnewline
\hline
$\Kinput$ & 300  & number of neurons  each input source is connected to (out-degree)\tabularnewline
\hline
\end{tabular}
\begin{tabular}{|@{\hspace*{1mm}}p{3cm}@{}|@{\hspace*{1mm}}p{4cm}@{}|@{\hspace*{1mm}}p{8.1cm}|}
\hline 
\multicolumn{3}{|>{\columncolor{gray}}c|}{\textbf{Simulation}}\tabularnewline
\hline 
\textbf{Name} & \textbf{Value } & \textbf{Description}\tabularnewline
\hline 
$T$ & $0.4$, $10$ or $2000\seconds$ & total simulation time \tabularnewline
\hline 
$\dtsim$ & $0.1\ms$ & time resolution \tabularnewline
\hline
$M$ & $10$ & number of random network realizations per parameter configuration \tabularnewline
\hline 
\end{tabular}
\caption{Network and simulation parameters}
\label{tab:parameters_1}
\end{table}

\begin{table}[H]
\renewcommand{\arraystretch}{1.2}
\begin{tabular}{|@{\hspace*{1mm}}p{3cm}@{}|@{\hspace*{1mm}}p{4cm}@{}|@{\hspace*{1mm}}p{8.1cm}|}
\hline 
\multicolumn{3}{|>{\columncolor{gray}}c|}{\textbf{Spike-train statistics}}\tabularnewline
\hline 
\textbf{Name} & \textbf{Value } & \textbf{Description}\tabularnewline
\hline 
$b$ & $10\ms$ & binsize for evaluation of Fano factor\tabularnewline
\hline 
\end{tabular}
\begin{tabular}{|@{\hspace*{1mm}}p{3cm}@{}|@{\hspace*{1mm}}p{4cm}@{}|@{\hspace*{1mm}}p{8.1cm}|}
\hline 
\multicolumn{3}{|>{\columncolor{gray}}c|}{\textbf{Perturbation sensitivity}}\tabularnewline
\hline 
\textbf{Name} & \textbf{Value } & \textbf{Description}\tabularnewline
\hline 
$t^*$ & $400\ms$ & perturbation time\tabularnewline
\hline 
$\dtpert$ & $0.5\ms$ & perturbation magnitude (time shift of one input spike)\tabularnewline
\hline
$\tobs$ & $10\seconds$ & observation time \tabularnewline
\hline
$\dtfilter$  & $1\ms$ & time resolution of low-pass filtered spiking activity \tabularnewline
\hline
$\taufilter$ & $20\ms$ & time constant of low-pass filter $h(t)$
\tabularnewline
\hline 
\end{tabular}
\caption{Parameters for evaluation of spike-train statistics and perturbation sensitivity}
\label{tab:parameters_2}
\end{table}

\subsection{Canceling of the synaptic-weight variance by the input variance}
\label{sec:Canceling of the synaptic-weight variance}
\subsection{Unspecific synapse loss and homeostasis}
\label{sec:unspecific_synapse_loss_with_homeostasis}

In this section, we expand our analysis of the linearized network dynamics towards a network in which all types of synapses ($\EE$,EI,IE,II) are removed. 
Accordingly, the homeostatic upscaling affects all types of synapses (EE,EI,IE,II) such that the target firing rate is reached by applying
the same factor $c$ to all synaptic weights and the initial proportion of 
the different synapse types is kept constant ($c\cdot\JEE=c\cdot\JIE=c\cdot\J$ and   $c\cdot\JEI=c\cdot\JII=-cg\cdot\J$ with $c\geq1$).

\begin{figure}[ht!]
  \centering
  \includegraphics[width=\textwidth]{\figdir/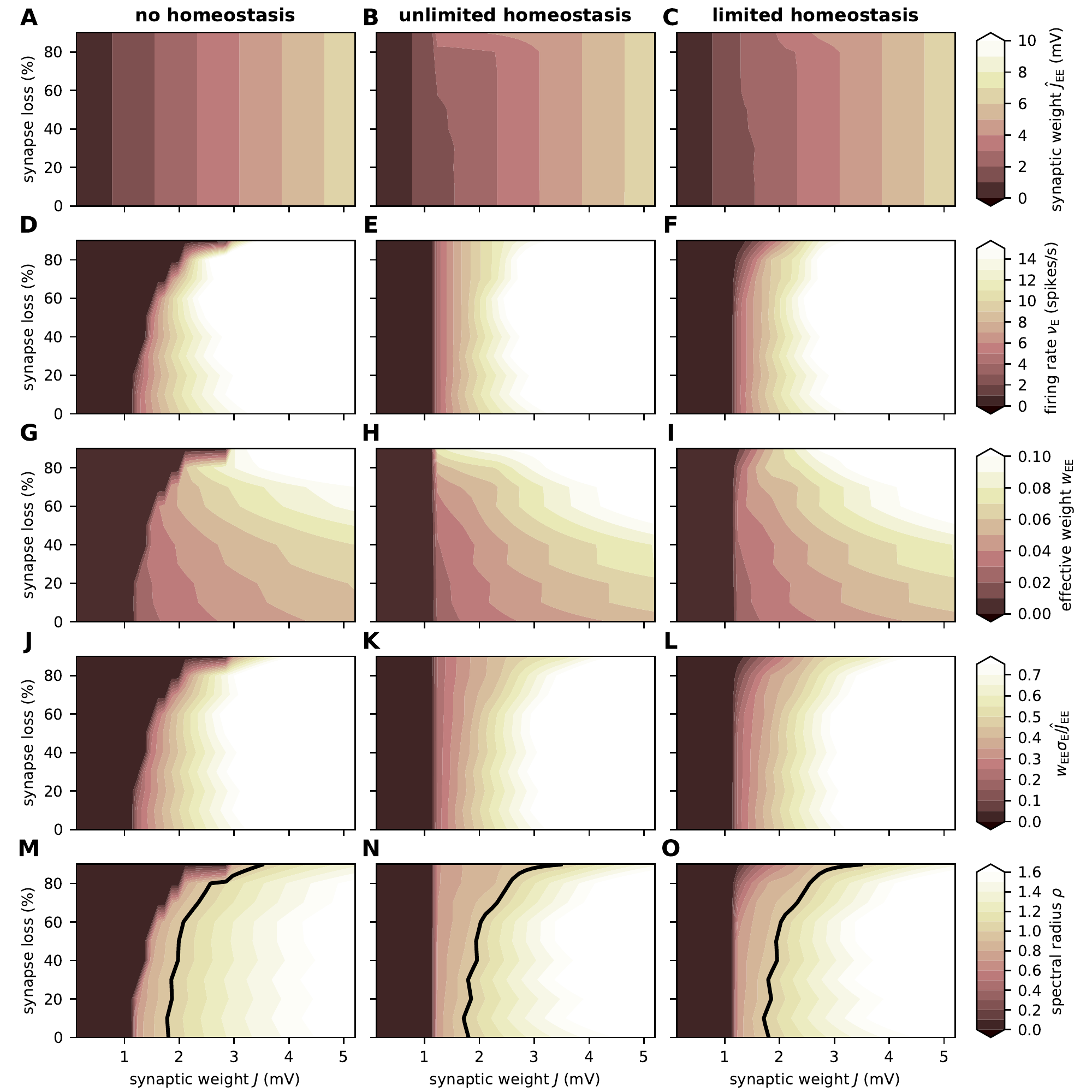}
  \caption{%
    \textbf{Mean-field theory applied to network with unspecific synapse loss and unspecific synaptic upscaling.}
    Dependence of
    the synaptic weight $\JfbEE$ (\textbf{A}--\textbf{C}),
    the average firing rate $\nuE$ of the excitatory population (\textbf{D}--\textbf{F}),
    the effective weight $\wEE$ of $\EE$ connections (\textbf{G}--\textbf{I}), 
    the ratio $\wEE\sigmaE/\JfbEE$ (\textbf{J}--\textbf{L}), 
    and the spectral radius $\rho$ (\textbf{M}--\textbf{O}) 
    on the synaptic weight $J$ and the degree of synapse loss in the absence of homeostatic compensation (left column), as well as with unlimited (middle column) and limited firing rate homeostasis (right column).
    Superimposed black curves in (M--O) mark instability lines $\rho=1$.
    Same parameters as in network simulations (see \cref{sec:suppl_network_model} and \cref{sec:suppl_parameters}).
    }
  \label{fig:theory_unspecific_synapse_loss}
\end{figure}

We observe that synapse-unspecific network dilution leads to a drop in firing rate (\cref{fig:theory_unspecific_synapse_loss} D), but this drop is not as pronounced as if only EE synapses are removed  (\cref{fig:theory} D).
For small and moderate degrees of synapse loss, the firing rate changes only little. Upscaling $\J$ compensates for this and fully restores the firing rates (\cref{fig:theory_unspecific_synapse_loss} E), even for high levels of synapse loss.
In the absence of homeostasis, synapse unspecific network dilution reduces the spectral radius (\cref{fig:theory_unspecific_synapse_loss} M), but this effect is weaker as if only EE synapses were removed (\cref{fig:theory} M). For small and moderate degrees of synapse loss, the spectral radius is hardly affected.
Upscaling $\J$ fully recovers the spectral radius in the stable regime ($\rho<1$). Close to the transition from stable to unstable ($\rho=1$, black contour line), recovery of the spectral radius is approximately achieved (\cref{fig:theory_unspecific_synapse_loss} N).

\end{document}